%% file: ms.tex
\documentclass[10pt,conference,english]{IEEEtran}
\IEEEoverridecommandlockouts
\usepackage[centertags]{amsmath}
\usepackage{amsfonts}
\usepackage{amssymb}
\usepackage{epsfig}
\usepackage{amsmath,  amssymb}
\usepackage{graphicx}
\usepackage[centertags]{amsmath}
\usepackage{clrscode}
\usepackage{cite}
\usepackage{fullpage}
\usepackage{amsmath,amssymb,mathrsfs,pseudocode}
\usepackage{color}
\usepackage[normalem]{ulem}
\usepackage{psfrag}
\usepackage{url}
\usepackage{babel}
\usepackage{authblk}
\usepackage{mathtools}
\usepackage[linesnumbered,ruled]{algorithm2e}
\usepackage[left=54pt, right=54pt, top=51pt, bottom=58pt]{geometry}
\usepackage{xspace}% essential for newcommand endings
\usepackage{multicol}
\usepackage{booktabs} % nice table objects
\usepackage{multirow}
\usepackage[x11names,dvipsnames,table]{xcolor}

\usepackage{subfigure}

\usepackage{stackrel}
\usepackage{extarrows}

\newtheorem{theorem}{Theorem}%[section]
\newtheorem{lemma}{Lemma}

\newtheorem{proposition}{Proposition}%[section]
\newtheorem{definition}{Definition}%[section]
\newtheorem{corollary}{Corollary}%[section]
\newtheorem{remark}{Remark}%[section]

\include{defns}  % default definitions
\newcommand{\parastoo}{\textcolor{black}}

\newcommand{\Roy}{\textcolor{red}}

\newcommand{\Royy}{\textcolor{black}}

\newcommand{\Roycr}{\textcolor{black}}

\newcommand{\Rarxiv}{\textcolor{black}}

\allowdisplaybreaks
\begin{document}

\title{Information Leakage in Zero-Error Source Coding: A Graph-Theoretic Perspective}

\author{Yucheng Liu$^{\dag}$\thanks{This work was supported by the ARC Discovery Scheme DP190100770; the US National Science Foundation Grant CNS-1815322; and the ARC Future Fellowship FT190100429.}, Lawrence Ong$^{\dag}$, Sarah Johnson$^{\dag}$, Joerg Kliewer$^{*}$, Parastoo Sadeghi$^{\ddag}$, and Phee Lep Yeoh$^{\S}$\\\vspace{-0mm}
$^{\dag}$The University of Newcastle, Australia (emails: \{yucheng.liu, \hspace{-0.5mm}lawrence.ong, \hspace{-0.5mm}sarah.johnson\}@newcastle.edu.au)\\
$^{*}$New Jersey Institute of Technology, USA (email: jkliewer@njit.edu)\\
$^{\ddag}$University of New South Wales, Australia (email: p.sadeghi@unsw.edu.au)\\
$^{\S}$University of Sydney, Australia (email: phee.yeoh@sydney.edu.au)
}

%\footnote{The work of P. Sadeghi was supported by the ARC Future Fellowship, FT190100429.}

%Emails:  $^{\dag}$\{yucheng.liu, parastoo.sadeghi\}@anu.edu.au, $^{*}$\{ni.ding, thierry.rakotoarivelo\}@data61.csiro.au}

%\author{}

%\author{Parastoo Sadeghi$^{\dag}$, Fatemeh Arbabjolfaei$^{*}$, and Young-Han Kim$^{*}$\\\vspace{-0mm}
%$^{\dag}$Research School of Engineering, Australian National University, Canberra, ACT, 2601, Australia\\
%$^{*}$Department of Electrical and Computer Engineering, University of California, San Diego, CA 92093, USA\\
%Emails:  parastoo.sadeghi@anu.edu.au, \{farbabjo, yhk\}@ucsd.edu}

\maketitle

%\footnote{The work of P. Sadeghi was supported by the ARC Future Fellowship, FT190100429.}

\input{abstract_isit_2021.tex}

\input{introduction_isit_2021.tex}
\input{model_isit_2021.tex}

\input{main_results_isit_2021.tex}

\input{proofs_isit_2021.tex}

\bibliographystyle{IEEEtran}
\bibliography{references}

\end{document}

%% file: defns.tex
\usepackage{xspace}
\usepackage{bbm}
\input{mathlig}

\newcommand{\muspace}{\mspace{1mu}}

\DeclareRobustCommand{\scond}{\mathchoice{\muspace\vert\muspace}{\vert}{\vert}{\vert}}
\mathlig{|}{\scond}

\DeclareRobustCommand{\discint}{\mathchoice{\mspace{-1.5mu}:\mspace{-1.5mu}}{\mspace{-1.5mu}:\mspace{-1.5mu}}{:}{:}}
\mathlig{::}{\discint}

\newcommand{\Ec}{\mathcal{E}}

\newcommand{\Hc}{\mathcal{H}}
\newcommand{\Ic}{\mathcal{I}}

\newcommand{\Kc}{\mathcal{K}}
\newcommand{\Lc}{\mathcal{L}}
\newcommand{\Mc}{\mathcal{M}}

\newcommand{\Rc}{\mathcal{R}}
\newcommand{\Sc}{\mathcal{S}}
\newcommand{\Tc}{\mathcal{T}}

\newcommand{\Xc}{\mathcal{X}}
\newcommand{\Yc}{\mathcal{Y}}
\newcommand{\Zc}{\mathcal{Z}}

\newcommand{\Kcal}{\mathcal{K}}

\newcommand{\Pcal}{\mathcal{P}}

\newcommand{\Av}{{\bf A}}

\newcommand{\Xv}{{X}}

\newcommand{\Wv}{{\bf W}}

\newcommand{\dv}{{\bf d}}

\newcommand{\xv}{{x}}

\newcommand{\Yt}{{\tilde{Y}}}

\newcommand{\xt}{{\tilde{x}}}

\def\d{\delta}

\def\textiid{i.i.d.\@\xspace}
\newcommand\iid{\ifmmode\text{ i.i.d. } \else \textiid \fi}

\def\clap#1{\hbox to 0pt{\hss#1\hss}}
\def\mathclap{\mathpalette\mathclapinternal}
\def\mathclapinternal#1#2{%
  \clap{$\mathsurround=0pt#1{#2}$}}

\let\oldstackrel\stackrel
\renewcommand{\stackrel}[2]{\oldstackrel{\mathclap{#1}}{#2}}

%% file: abstract_isit_2021.tex
\begin{abstract}

We study the information leakage to a guessing adversary in zero-error source coding. The source coding problem is defined by a confusion graph capturing the distinguishability between source symbols. The information leakage is measured by the ratio of the adversary's successful guessing probability after and before eavesdropping the codeword, maximized over all possible source distributions. Such measurement under the basic adversarial model where the adversary makes a single guess and allows no distortion between its estimator and the true sequence is known as the maximum min-entropy leakage or the maximal leakage in the literature. 
We develop a single-letter characterization of the optimal normalized leakage under the basic adversarial model, together with an optimum-achieving scalar stochastic mapping scheme. An interesting observation is that the optimal normalized leakage is equal to the optimal compression rate with fixed-length source codes, both of which can be simultaneously achieved by some deterministic coding schemes. We then extend the leakage measurement to generalized adversarial models where the adversary makes multiple guesses and allows certain level of distortion, for which we derive single-letter lower and upper bounds.

%Considering a basic adversarial model where the adversary makes a single guess and allows no distortion between its estimator and the true sequence, the leakage has been analyzed is known as the maximum min-entropy leakage (Braun \emph{et al.}, 2009), and the maximal leakage (Issa \emph{et al.}, 2016). We first develop a single-letter characterization of the optimal normalized leakage for the basic adversarial model, together with an optimum-achieving scalar mapping scheme. We then extend the leakage measurement to generalized adversarial models where the adversary makes multiple guesses and allows certain level of distortion, for which we derive single-letter lower and upper bounds. 

\end{abstract}

%% file: introduction_isit_2021.tex
\section{Introduction}\label{sec:intro}

We study the fundamental limits of information leakage in zero-error source coding from a graph-theoretic perspective. 
%This work studies the information leakage in the zero-error source coding problem, where a loselossly compressed information source is transmitted to a legitimate receiver and eavesdropped by a guessing adversary. 
%

\begin{figure}[ht]
\begin{center}
\subfigure[][]{
\label{fig:toy:exm:receiver}
\includegraphics[scale=0.375]{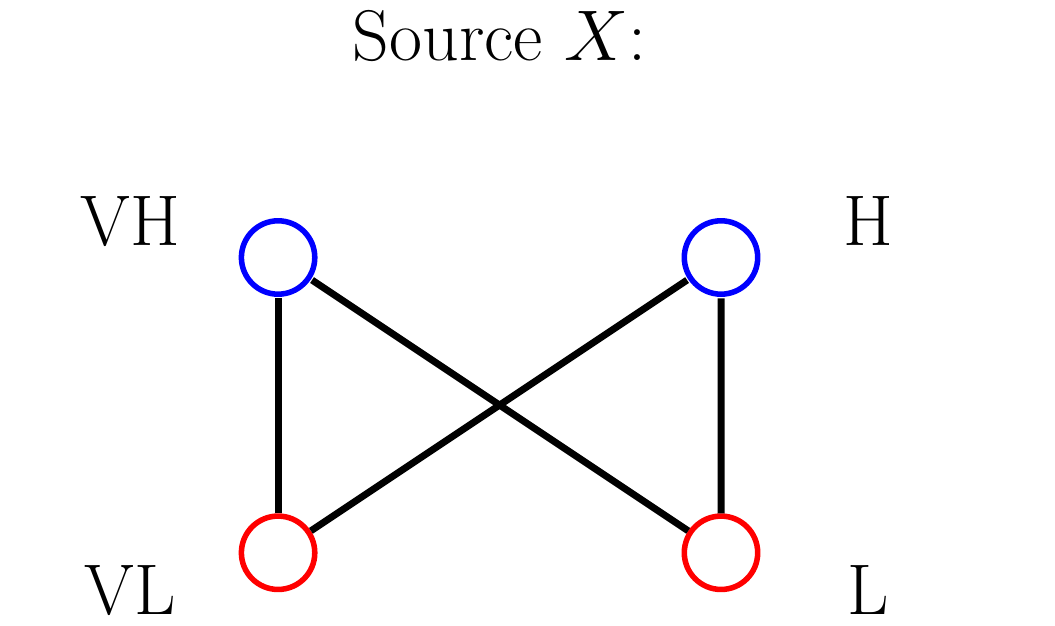}}
\subfigure[][]{
\label{fig:toy:exm:adversary}
\includegraphics[scale=0.3625]{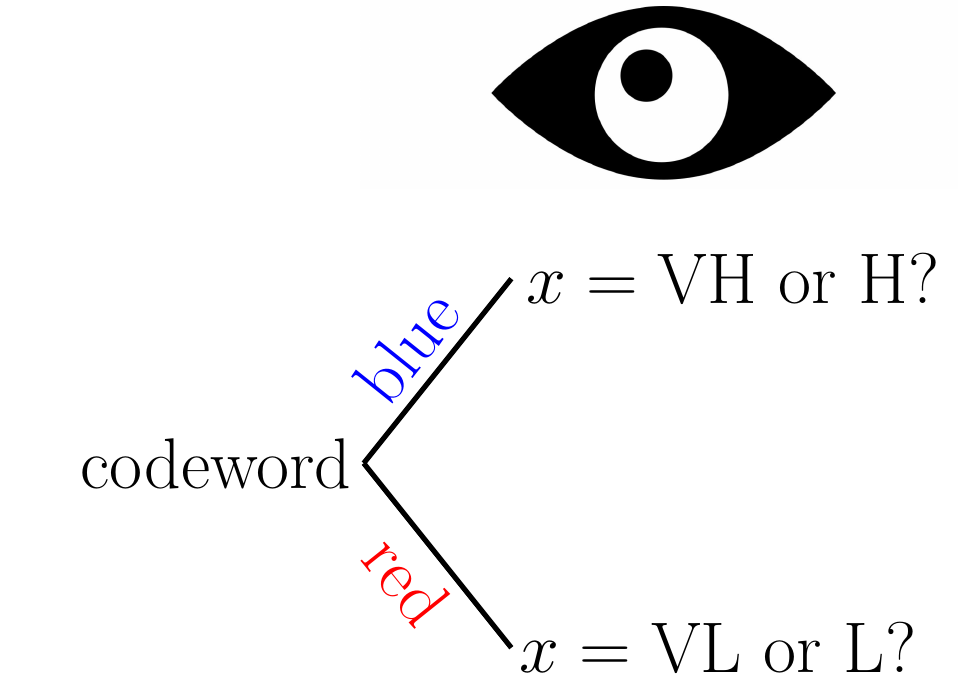}}
\caption{(a) From the supervisor's perspective, symbols ${\rm VH}$ (very high) and ${\rm H}$ (high) are indistinguishable (i.e., need not to be distinguished), and so are symbols ${\rm VL}$ (very low) and ${\rm L}$ (low). We draw an edge between any two distinguishable symbols, and then to satisfy the supervisor, we can only map non-adjacent symbols to the same codeword. 
%any adjacent symbols must not be mapped to the same codeword. 
(b) An adversary eavesdrops the codeword, based upon which it guesses the exact water level.\vspace{-8mm}}
\label{fig:toy:exm}
\end{center}
\end{figure}

%Source coding considers efficient compression of data produced by an information source, which is one of the most basic topics in information theory since its introduction by Shannon \cite{shannon1948mathematical}.  
Source coding \cite{shannon1948mathematical} considers compression of an information source to represent data with fewer number of bits by mapping multiple source sequences to the same codeword.
%
%Source coding or compression is one of the most basic and classic topic in information theory ever since its introduction by Shannon. 
%
%Zero-error
%
Suppose we observe a source $X$ and wish to transmit a compressed version of the source to a legitimate receiver. 
%without error, from whose perspective \Royy{some source symbols are to be distinguished and some are not. We say two source symbols/sequences are distinguishable if they are to be distinguished by the receiver}. 
From the receiver's perspective, some source symbols are to be distinguished and some are not. We say two source symbols/sequences are distinguishable if they are to be distinguished by the receiver. 
%
%the source symbols are not all distinguishable (i.e., the receiver does not need to differentiate between all the source symbols for X). 
%To achieve zero decoding error, any distinguishable source sequences must \emph{not} be mapped to the same codeword. 
For successful decoding, any distinguishable source sequences \emph{must not} be mapped to the same codeword. 
The distinguishability relationship is characterized by the \emph{confusion graph} $\Gamma$ for the source. 
Such graph-theoretic model has various applications in the real world. 
Consider the toy example in Figure \ref{fig:toy:exm:receiver}, where $X$ denotes the water level of a reservoir, %or a storage tank in some industrial plants. 
%Suppose that a supervisor only needs to know whether the water level is high or low such that if it is low then some actions are to be taken. 
and a supervisor only needs to know whether the water level is relatively high or low to determine whether a refilling is needed.

The source coding model we consider was originally introduced by K\"orner \cite{korner1973coding}, 
where a vanishing error probability is allowed 
%a small fraction of source sequences is allowed to be ``improperly" compressed (i.e., confusable sequences being mapped to the same codeword) 
and the resulted optimal compression rate is defined as the \emph{graph entropy} of the confusion graph. 
%Similar zero-error setup has been investigated in \cite{ddd} where the information ratio (definition) is studied. 
%Also, the joint source-channel coding problem was studied \Royy{by Wang and Shayevitz \cite{wang2017graph}} in which the source coding part considers the zero-error graph-theoretic setup same as ours. 
More recently, Wang and Shayevitz \cite{wang2017graph} analyzed the joint source-channel coding problem based on the same zero-error graph-theoretic setting as our model for the source coding.

Suppose that the transmitted codeword is eavesdropped by a guessing adversary, who knows the source distribution $P_X$ and tries to guess the true source sequence via maximum likelihood estimation within a certain number of trials. 
See Figure \ref{fig:toy:exm:adversary} for an example. 
Before observing the codeword, the adversary will guess the most likely water level among all four levels. After observing the codeword, say ``blue", it will guess the more likely water level between ${\rm VH}$ and $\rm H$.
%Before observing the codeword, the adversary will guess the most likely water level among all four levels. After observing the codeword, say "blue", it will guess the more likely water level between ${\rm VH}$ and ${\rm H}$.
%\Roy{For an instance, suppose an adversary tries to guesses the exact water level of the reservoir described in Figure \ref{fig:toy:exm}}. 
%That is, the adversary aims at maximizing its probability of successfully guessing the source sequence within a certain number of trials. 
%Compared with guessing the source blindly, that is, guessing based on only the source distribution and without observation of the transmitted message, the probability of 
Compared with guessing blindly (i.e., based only on $P_X$), the average successful guessing probability will increase as the adversary eavesdrops the codeword. 
We measure the information leakage from the codeword to the adversary by such a probability increase. 
%\Roy{For an instance, again consider the toy example in Figure \ref{fig:toy:exm}. Suppose an adversary tries to guess the exact water level of the reservoir}.
More specifically, the leakage is quantified as the ratio between the adversary's probability of successful guessing \emph{after and before} observing the codeword. 
This way of measuring information leakage was originally introduced by Smith \cite{smith2009foundations}, leading to the leakage metric commonly referred to as min-entropy leakage.

Quite often in practice, the compression scheme is designed without knowing the exact source distribution. In such case, one can consider the \emph{worst-case} leakage, which is the information leakage maximized over all possible source distributions $P_X$ over the fixed alphabet $\Xc$. 
The worst-case variant of min-entropy leakage, namely the maximum min-entropy leakage, was developed by Braun \emph{et al.} \cite{braun2009quantitative}.

\iffalse
The basic idea of measuring leakage by the increase in the adversary's successful probability of guessing some $X$ was introduced \Royy{by Smith} \cite{smith2009foundations}, leading to the definition of the privacy metric known as \emph{min-entropy leakage}. 
%for an arbitrary but fixed $P_X$,
%The idea was further explored in \cite{braun2009quantitative} where the
A worst-case variant of the min-entropy leakage, namely the \emph{maximum} min-entropy leakage, was developed later \Royy{by Braun \emph{et al.}} \cite{braun2009quantitative}. 
Note that the adversary in \cite{smith2009foundations,braun2009quantitative} makes only a single guess and allows no distortion between its estimator and the true value. 
\fi

\Royy{A similar idea} was independently explored by Issa \emph{et al.} \cite{issa2016operational,issa2019operational} in a different setup where the adversary is interested in guessing some randomized function $U$ of $X$ rather than $X$ itself. 
The worst-case metric under such scenario is named as the \emph{maximal leakage}. 
Interestingly, despite their different operational meanings, the maximal leakage and maximum min-entropy leakage turn out to be equal. 
%Interestingly, despite the different operational meanings behind the maximal leakage and maximum min-entropy leakage, they turn out to have the same computable expression and thus equal. 
% 
\iffalse
The basic idea of such measurement of information leakage was originally introduced in \cite{smith2009foundations} for an arbitrary but fixed $P_X$, and further explored in \cite{braun2009quantitative} where the worst-case scenario was considered. 
The worst-case metric defined in \cite{braun2009quantitative} allows the adversary to make one guess only and is commonly referred to as the \emph{maximum min-entropy leakage} in the literature afterwards. 
\fi
%
%Another related work is \Royy{by Shkel and Poor} \cite{shkel2020compression}, where the leakage of sensitive information in compression systems has been studied under multiple leakage metrics, including the maximal leakage, and the assumption that the source code must be deterministic yet a random secret key is shared between the sender and the receiver. 
For more works studying the maximal leakage or the maximum min-entropy leakage and their variants from both the information-theoretic and computer science perspectives, see \cite{liao2017hypothesis,karmoose2019privacy,esposito2019learning,yuchengliu2020isit,zhou2020weakly,wu2020optimal,m2012measuring,espinoza2013min,alvim2014additive,smith2015recent}. 
In another related work by Shkel and Poor \cite{shkel2020compression}, leakage in compression systems has been studied considering multiple leakage metrics, including the maximal leakage, under the assumption that the source code must be deterministic yet a random secret key is shared between the sender and the receiver. 
%Multiple leakage metrics were considered in \cite{shkel2020compression}, including the maximal leakage. 
%Another related work is \cite{shkel2020compression}, where the leakage of sensitive information in compression systems has been studied under the assumption that the source code must be deterministic yet a random secret key is shared between the sender and the receiver. 
%Multiple leakage metrics were considered in \cite{shkel2020compression}, including the maximal leakage. 

Clearly we wish to keep the information leakage as small as possible by \Royy{smartly designing a (possibly stochastic) source coding scheme}. % while ensuring zero-error decoding at the legitimate receiver. 
%Thus to mitigate the information leakage due to the exposure of the message, we want to keep such ratio as small as possible. 
%Therefore, we ask the following two fundamental questions: what is the optimal (i.e., minimum) leakage rate that can be achieved asymptotically while ensuring zero-error decoding at the legitimate receiver? 
Therefore, our fundamental objective is to characterize the minimum leakage (normalized to the source sequence length) under the zero-error decoding requirement %(normalized over the source sequence length) 
and also the optimum-achieving mapping scheme. 

{\it Contributions and organization:}
%\Royy{We formally detail our problem in Section \ref{sec:model}. 
In Section II, we detail the problem of information leakage in source coding. In particular, we start with the basic adversarial model where the adversary makes a single guess and allows no distortion\footnote{When no distortion is allowed, the adversary must guess the actual source sequence to be considered successful.}, and thus the resulting privacy metric is the normalized version\footnote{The normalized version is appropriate as we compress a source sequence.} of the maximal leakage \cite{issa2019operational} or the normalized maximum min-entropy leakage \cite{braun2009quantitative}. 
Our main contributions are as follows: 
%\begin{enumerate}

1) In Section \ref{sec:maxl}, we develop a single-letter characterization for the optimal normalized maximal leakage for the basic adversarial model. We also design a scalar stochastic mapping scheme that achieves this optimum. 
An interesting observation is that the optimal leakage can also be achieved using deterministic codes that simultaneously achieve the optimal fixed-length zero-error compression rate. 
%An interesting observation is that the optimal leakage can in fact be simultaneously achieved with the optimal fixed-length zero-error compression rate by deterministic mapping. 
%of the optimal (normalized) maximum min-entropy leakage (maximal leakage). %within the model described in Section \ref{sec:model}. 

2) In Section \ref{sec:extensions}, we extend our adversarial model to allow multiple guesses and distortion between an estimator (guess) and the true sequence. 
%in two directions: the adversary can make multiple guesses and the adversary allows certain level of distortion between its estimator and the true sequence. 
Inspired by the notion of confusion graphs, we characterize the relationship between a sequence and its acceptable estimators by another graph defined on the source alphabet, resulting in a novel leakage \Royy{measurement}. 

3) We then show that the optimal normalized leakage under the generalized models is always upper-bounded by the result in the original setup. Single-letter lower bounds (i.e., converse results) are also established. 
%\end{enumerate}

%Some graph-theoretic notions are reviewed in Appendix \ref{app:basic:graph:theoretic:notions}. 
%A brief review on basic graph-theoretic definitions is given in Appendix \ref{app:basic:graph:theoretic:notions}. 
We also include a brief review on basic graph-theoretic definitions in Appendix \ref{app:basic:graph:theoretic:notions}. 

{\it Notation:}
For non-negative integers $a$ and $b$, $[a]$ denotes the set $\{1,2,\dots,a\}$, and $[a:b]$ denotes the set $\{a,a+1,\dots,b\}$. If $a>b$, $[a:b]=\emptyset$. 
For a finite set $A$, $|A|$ denotes its cardinality. %, and $$
For two sets $A$ and $B$, %$A\setminus B=\{ i\in A:i\notin B \}$, and 
$A\times B$ denotes their Cartesian product. 
For a sequence of sets $A_1,A_2,\dots,A_t$, we may simply use $\prod_{j\in [t]}A_j$ to denote their Cartesian product. 
For any discrete random variable $Z$ with probability distribution $P_Z$, we denote its alphabet by $\Zc$ with realizations $z\in \Zc$. 
For any $K\subseteq \Zc$, $P_Z(K)\doteq \sum_{z\in K} P_Z(z)$. 

%% file: model_isit_2021.tex
\section{System Model and Problem Formulation}\label{sec:model}

{\textbf{Source coding with confusion graph $\Gamma$:}}
Consider a discrete memoryless stationary information source $X$ that takes values in the alphabet $\Xc$ with full support. 
We wish to stochastically compress a source sequence $X^t\doteq (X_1,X_2,\dots,X_t)$ to some codeword $Y$ that takes values in the code alphabet $\Yc$ and transmit it to a legitimate receiver via a noiseless channel.  
The randomized mapping scheme from $\Xc$ to $\Yc$ is denoted by the conditional distribution $P_{Y|X^t}$. 
%The mapping scheme from $\Xc$ to $\Yc$ is denoted by the conditional probability distribution $P_{Y|X^t}$. 

%From the perspective of the legitimate receiver, any two source values are either \emph{confusable} or \emph{distinguishable}. 
To the receiver, the distinguishability relationship among source symbols is characterized by a confusion graph $\Gamma$, where the vertex set is the source alphabet, i.e., $V(\Gamma)=\Xc$, and any two symbols $x,x'\in \Xc$ are adjacent in $\Gamma$, i.e., $\{x,x'\}\in \Ec(\Gamma)$, iff they are distinguishable with each other. 
%any two source values are distinguishable if they should be BLA otherwise indistinguishable. 
%
%Such relationship is characterized by the confusion graph $\Gamma$, which is an undirected graph with vertex set equals to the source alphabet, i.e., $V(\Gamma)=\Xc$. 
%Any two source values $x,x'\in \Xc$ are connected in $\Gamma$, i.e., $\{x,x'\}\in \Ec(\Gamma)$, if and only if (iff) they are confusable with each other. 
%A brief review on basic graph-theoretic definitions is given in Appendix \ref{app:basic:graph:theoretic:notions}. 
%
Any two source sequences, $x^t=(x_1,\dots,x_t)\in \Xc^t$ and $v^t=(v_1,\dots,v_t)\in \Xc^t$, are distinguishable iff at some $j\in [t]$, $x_j$ and $v_j$ are distinguishable. 
Therefore, the distinguishability among source sequences of length $t$ is characterized by the confusion graph $\Gamma_t$, which is defined as the $t$-th power of $\Gamma$ with respect to the \emph{OR (disjunctive) graph product} \cite[Section 3.4]{scheinerman2011fractional}: $\Gamma_t=\Gamma \lor \Gamma \lor \dots \lor \Gamma=\Gamma^{\lor t}$. 
%We recall the OR graph product as follows. 
%%
%\begin{definition}[OR product,\cite{scheinerman2011fractional}]  \label{def:or:product}
%Given two undirected graphs $G_1,G_2$, the OR product $G=G_1\lor G_2$ is constructed as $V(G)=V(G_1)\times V(G_2)$ and $\{ (i_1,i_2),(j_1,j_2) \}\in \Ec(G)$ iff $\{ i_1,j_1 \}\in \Ec(G_1)$ or $\{ i_2,j_2 \}\in \Ec(G_2)$. 
%\end{definition}

To ensure zero-error decoding, any two source sequences that can be potentially mapped to the same codeword must not be distinguishable. More formally, given some $P_{Y|X^t}$, let 
\begin{align}
\Xc^t_{P_{Y|X^t}}(y)\doteq \{ x^t\in \Xc^t:P_{Y|X^t}(y|x^t)>0 \}  \label{eq:notation:conditional:support}
\end{align}
denote the set of all $x^t$ mapped to $y$ with nonzero probability. 
When there is no ambiguity, we simply denote $\Xc^t_{P_{Y|X^t}}(y)$ by $\Xc^t(y)$. 
Therefore, a mapping scheme $P_{Y|X^t}$ is \emph{valid} iff %it satisfies the following compression requirement: %to ensure zero-error decoding at the legitimate receiver:  
\begin{align}
\Xc^t(y)\in \Ic(\Gamma_t), \quad \forall y\in \Yc,  \label{eq:confusion:requirement}
\end{align}
where $\Ic(\cdot)$ denotes the set of independent sets of a graph \Rarxiv{(cf. Appendix \ref{app:basic:graph:theoretic:notions})}.

{\textbf{Leakage to a guessing adversary:}}
As a starting point, we assume that the adversary makes a single guess after observing each codeword and allows no distortion between its estimator sequence and the true source sequence. 
%which is defined as follows. 
%In other words, the adversary always guesses the most probable source sequence. 
%The privacy leakage is defined as the ratio between the expected probabilities of the most probable source sequences 
%\emph{after and before} observing the transmitted codeword. 
%The leakage is maximized over all possible source distributions $P_X$ over the fixed alphabet $\Xc$. %, and then normalized over the source sequence length. 

%Consider any source coding problem\footnote{When there is no ambiguity, to refer to as a zero-error source coding problem with confusion graph $\Gamma$, we just say a source coding problem $\Gamma$. } $\Gamma$. 
%The maximal leakage\footnote{Note that we have adopted the name of maximal leakage from \cite{issa2019operational}, which equals to the maximum min-entropy leakage from \cite{braun2009quantitative}. } for a given sequence length $t$ and a given valid mapping $P_{Y|X^t}$, denoted by $\Lc(\Gamma,t,P_{Y|X^t})$, is defined as 

Consider any source coding problem $\Gamma$.\footnote{When there is no ambiguity, instead of saying a zero-error source coding problem with confusion graph $\Gamma$, we just say a source coding problem $\Gamma$.}
The maximal leakage\footnote{Note that we have adopted the name of maximal leakage \cite{issa2019operational}, which is equivalent to the maximum min-entropy leakage \cite{braun2009quantitative}. } for a given sequence length $t$ and a given valid mapping $P_{Y|X^t}$ is defined as follows:\footnote{For notation brevity, we drop the reference to $\Gamma$ noting that all leakage measures defined in this paper are dependent on $\Gamma$. }
\begin{align}
L^t(P_{Y|X^t})
&\doteq \log \max_{P_X} \frac{ {\mathbb E}_{Y}\left[\max\limits_{x^t\in \Xc^t} P_{X^t|Y}(x^t|Y)\right] }{ \max\limits_{x^t\in \Xc^t} P_{X^t}(x^t) }  \label{eq:def:maxl:leakage:rate:any:t:mapping:operational}  \\
&=\log \max_{P_X} \frac{\sum\limits_{y\in \Yc} \max\limits_{x^t\in \Xc^t} P_{X^t,Y}(x^t,y)}{\max\limits_{x^t\in \Xc^t}P_{X^t}(x^t)}  \label{eq:def:max:leakage:rate:any:t:mapping}  \\
&=\log \sum\limits_{y\in \Yc} \max\limits_{x^t\in \Xc^t} P_{Y|X^t}(y|x^t),\label{eq:def:max:leakage:rate:simplified:any:t:mapping}
\end{align}
where $\eqref{eq:def:max:leakage:rate:simplified:any:t:mapping}$ follows from \cite[Proposition 5.1]{braun2009quantitative}. %, which is equal to the Sibson mutual information of order infinity \cite{verdu2015alpha} betwen $X^t$ and $Y$. %, $I_{\infty}^{\rm Sibson}(X^t;Y)$ . 
%\Roy{Change the notation for $\Lc(\Gamma,t,P_{Y|X^t})$ to $L(\Gamma,t,P_{Y|X^t})$ to show the difference between it and $\Lc(\Gamma)$? }
%
%
The \emph{optimal} maximal leakage for a given $t$ is then defined as 
\begin{align}
\Lc^t&\doteq \inf_{\substack{P_{Y|X^t}:\\ \Royy{\Xc^t(y)\in \Ic(\Gamma_t)},\forall y\in \Yc}} L^t(P_{Y|X^t})\label{eq:def:max:leakage:rate:any:t},  %\\
%&=\inf_{\substack{P_{Y|X^t}:\\ \Xc^t(y)\in \Ic_{\rm max}(\Gamma_t),\forall y\in \Yc}} \frac{1}{t}\log \sum_{y\in \Yc} \max_{x^t\in \Xc^t} P_{Y|X^t}(y|x^t).  \label{eq:def:max:leakage:rate:simplified:any:t}
\end{align}
based upon which we can define the (optimal) maximal leakage rate as 
\begin{align}
\Lc&\doteq \lim_{t\to \infty} t^{-1} \Lc^t. \label{eq:def:max:leakage:rate} 
%=\lim_{t\to \infty}\inf_{\substack{P_{Y|X^t}:\\ \Xc^t(y)\in \Ic_{\rm max}(\Gamma_t),\forall y\in \Yc}} \Lc(\Gamma,t,P_{Y|X^t}). \label{eq:def:max:leakage:rate}  %\\
%&=\lim_{t\to \infty}\inf_{\substack{P_{Y|X^t}:\\ \Xc^t(y)\in \Ic_{\rm max}(\Gamma_t),\forall y\in \Yc}} \frac{1}{t}\log \sum_{y\in \Yc} \max_{x^t\in \Xc^t} P_{Y|X^t}(y|x^t).  \label{eq:def:max:leakage:rate:simplified}
\end{align}

%% file: main_results_isit_2021.tex
\section{Maximal Leakage Rate: Characterization}  \label{sec:maxl}

In the following we present a single-letter characterization of the maximal leakage rate $\Lc$. 

%In the following we present our first main result, a single-letter characterization of the maximal leakage rate $\Lc^*(\Gamma)$. % for any given source coding problem $\Gamma$. 

%Consider any source coding problem $\Gamma$. The maximal leakage rate $\Lc^*(\Gamma)$ is characterized as follows. 

\begin{theorem}  \label{thm:maxl}
For any source coding problem $\Gamma$, %we have 
\begin{align}
\Lc=\log \chi_f(\Gamma),  \label{eq:thm:maxl}
\end{align} 
%where $\chi_f(\Gamma)$ is the fractional chromatic number of $\Gamma$. 
\Rarxiv{where $\chi(\cdot)$ denotes the fractional chromatic number of a graph (cf. Definition \ref{def:fractional:chromatic})}. 
\end{theorem}

%To prove the above theorem, we first show an important lemma as follows. 
%The proof of Theorem \ref{thm:maxl} relies on the following lemmas. 
To prove Theorem \ref{thm:maxl}, we introduce several useful lemmas. 

We first show that given any mapping scheme, 
%The lemma below indicates that 
``merging" any two codewords does not increase the leakage (as long as the generated mapping is still valid).

More precisely, consider any sequence length $t$ and any valid mapping $P_{Y|X^t}$ such that there exists some \emph{mergeable} codewords $y_1,y_2\in \Yc$, $y_1\neq y_2$, satisfying $\Xc^t(y_1)\cup \Xc^t(y_2)\subseteq T$ for some $T\in \Ic_{\rm max}(\Gamma_t)$, 
\Royy{where $\Ic_{\rm max}(\cdot)$ denotes the set of \emph{maximal} independent sets of a graph} \Rarxiv{(cf. Appendix \ref{app:basic:graph:theoretic:notions})}. 
%We call such two codewords $y^1,y^2$ \emph{mergeable}. 
Construct $P_{Y_{1,2}|X^t}$ by merging $y_1$ and $y_2$ to a new codeword $y_{1,2}\notin \Yc$. That is, $\Yc_{1,2}=(\Yc\setminus \{y_1,y_2\}) \cup \{y_{1,2}\}$, and for any $x^t\in \Xc^t$, 
\begin{align}
&P_{Y_{1,2}|X^t}(y|x^t)  \nonumber  \\
&=\begin{cases}
P_{Y|X^t}(y_1|x^t)+P_{Y|X^t}(y_2|x^t), \quad \enskip \text{if $y=y_{1,2}$},  \\
P_{Y|X^t}(y|x^t), \qquad \qquad \qquad \qquad \enskip \text{otherwise}.
\end{cases}
\end{align}
%\begin{align}
%&P_{Y_{1,2}|X^t}(y|x^t)  \nonumber  \\
%&=\begin{cases}
%P_{Y|X^t}(y|x^t), \qquad \qquad \qquad \enskip y\in \Yc\setminus \{y_1,y_2\}, \\
%P_{Y|X^t}(y_1|x^t)+P_{Y|X^t}(y_2|x^t), \quad \quad \enskip y=y_{1,2}.
%\end{cases}
%\end{align}
Then we have the following result. 

\begin{lemma}  \label{lem:maxl:merging}
%We have 
$L^t(P_{Y_{1,2}|X^t})\le L^t(P_{Y|X^t})$. 
%\begin{align}
%\Lc(\Gamma,t,P_{Y^{1,2}|X^t})\le \Lc(\Gamma,t,P_{Y|X^t}). 
%\end{align}
\end{lemma}

\begin{IEEEproof}
It suffices to show $\max_{x^t\in \Xc^t} P_{Y_{1,2}|X^t}(y_{1,2}|x^t)$ being no larger than $\sum_{y\in \{y_1,y_2\}} \max_{x^t\in \Xc^t} P_{Y|X^t}(y|x^t)$ as 
\begin{align*}
&\max_{x^t\in \Xc^t} P_{Y_{1,2}|X^t}(y_{1,2}|x^t)  \\
&=\max_{x^t\in \Xc^t} (P_{Y|X^t}(y_1|x^t) + P_{Y|X^t}(y_2|x^t))  \\
&\le \max_{x^t\in \Xc^t} P_{Y|X^t}(y_1|x^t) + \max_{x^t\in \Xc^t} P_{Y|X^t}(y_2|x^t)  \\
&=\sum_{y\in \{y_1,y_2\}} \max_{x^t\in \Xc^t} P_{Y|X^t}(y|x^t),
\end{align*}
which completes the proof of the lemma. 
\end{IEEEproof}

%As suggested by Lemma \ref{lem:maxl:merging}, we can always merge without enlarging the privacy leakage to the adversary, as long as the mapping after merging is still valid. 
As specified in \eqref{eq:confusion:requirement}, for a valid mapping scheme, every codeword $y$ should correspond to an independent set of the confusion graph $\Gamma$. 
As a consequence of Lemma \ref{lem:maxl:merging}, to characterize the optimal leakage, it suffices to consider only those mapping schemes for which all codewords $y$ correspond to \emph{distinct} maximal independent sets of $\Gamma$. 
%it suffices to consider only those mapping schemes for which every $y$ corresponds to a \emph{different }\emph{maximal} independent set of $\Gamma$. 

To formalize this observation, for any sequence length $t$, define the distortion function $d_t:\Xc^t\times \Ic_{\rm max}(\Gamma_t) \to \{0,1\}$ such that for any $x^t\in \Xc^t$, $T\in \Ic_{\rm max}(\Gamma_t)$, 
\begin{align}
d(x^t,y)=\begin{cases} 0, \quad x^t\in T, \\ 1, \quad x^t\notin T. \end{cases}
\end{align}
%Then the lemma below holds, whose proof can be found in the full version of this paper \cite{isit:2021:arxiv}. 
Then the lemma below holds, whose proof is presented in Appendix \ref{app:proof:lem:maxl:merging:consequence}.

\begin{lemma}  \label{lem:maxl:merging:consequence}
%Consider any sequence length $t$. 
To characterize $\Lc^t$ defined in \eqref{eq:def:max:leakage:rate:any:t}, it suffices to assume the mapping $P_{Y|X^t}$ to satisfy that $\Yc=\Ic_{\rm max}(\Gamma_t)$ and $d(X^t,Y)=0$ almost surely. 
Thus by \eqref{eq:def:max:leakage:rate:simplified:any:t:mapping}, we have
%In other words, $\Lc^*(\Gamma,t)$ is equivalent to the solution to the following optimization problem: 
\begin{align}
\Lc^t%&=\inf_{\substack{P_{Y|X^t}:\\ \Yc=\Ic_{\rm max}(\Gamma_t),d(X^t,Y)=0}} L^t(P_{Y|X^t})  \nonumber  \\
&=\inf_{\substack{P_{Y|X^t}:\\ \Yc=\Ic_{\rm max}(\Gamma_t), \\d(X^t,Y)=0}} \log \sum_{y\in \Yc} \max_{x^t\in \Xc^t} P_{Y|X^t}(y|x^t).  \label{eq:maxl:PUT:optimization}
\end{align}
%In other words, one can, without loss of generality, replace the constraint 
%$$P_{Y|X^t}:\Xc^t(y)\in \Ic(\Gamma_t),\forall y\in \Yc$$ 
%under infimum in \eqref{eq:def:max:leakage:rate:simplified} to 
%%$$P_{Y|X^t}: \Yc=\Ic_{\rm max}(\Gamma_t), X^t\in Y.$$ 
%$$P_{Y|X^t}: \Yc=\Ic_{\rm max}(\Gamma_t),d(X^t,Y)=0.$$
\end{lemma}

The solution\footnote{By solution we mean the optimal objective value of the problem.} to the optimization problem on the right hand side of \eqref{eq:maxl:PUT:optimization} in Lemma \ref{lem:maxl:merging:consequence} is characterized by \cite[Corollary 1]{Liao2019tunable}, 
based upon which we have the following result. 
%, which is re-stated in our notation as follows. 

%As suggested by Lemma \ref{lem:maxl:merging}, for any $t$, solving BLABLA is equivalent to solving the \emph{privacy-utility tradeoff} studied in \cite[(52), Section V]{Liao2019tunable} with the following \emph{hard distortion} constraint: BLABLA. 

%%For any sequence length $t$, define the distortion function $d:\Xc^t\times \Ic_{\rm max}(\Gamma_t) \to \{0,1\}$ such that for any $x^t\in \Xc^t$, $y\in \Ic_{\rm max}(\Gamma_t)$, 
%%\begin{align}
%%d(x^t,y)=\begin{cases} 0, \quad x^t\in y, \\ 1, \quad x^t\notin y. \end{cases}
%%\end{align}
%%According to Lemma \ref{lem:maxl:hard:distortion}, $\Lc(\Gamma,t)$ is equivalent to the solution to the following optimization problem: 
%%\begin{align}
%%\inf_{\substack{P_{Y|X^t}:\\ \Yc=\Ic_{\rm max}(\Gamma_t),d(X^t,Y)=0}} \frac{1}{t} \log \sum_{y\in \Yc} \max_{x^t\in \Xc^t} P_{Y|X^t}(y|x^t),  \label{eq:maxl:PUT:optimization}
%%\end{align}
%%%for some distortion function $d(\cdot,\cdot)$ defined as 
%%%\begin{align}
%%%d(x^t,y), \quad \forall x^t\in \Xc^t, y\in \Ic_{\rm max}(\Gamma_t)
%%%\end{align}
%%which is characterized by \cite[Corollary 1]{Liao2019tunable}, re-stated in our notation as follows. 
%%%The optimal tradeoff for such problem is summarized by the lemma below. 

\begin{lemma}  \label{lem:maxl:hard:distortion}
%[Corollary 1, \cite{Liao2019tunable}]  \label{lem:maxl:hard:distortion}
%The solution to the optimization problem \eqref{eq:maxl:PUT:optimization} is $-\frac{1}{t} \log \eta$, where $\eta$ is the solution to the following maximin problem:
%Consider any sequence length $t$.  
%We have 
$\Lc^t=-\log \eta,$ where $\eta$ is the solution to the following maximin problem: 
\begin{subequations}  \label{eq:maxl:hard:distortion:optimization}
\begin{align}
\text{maximize} \quad &\min_{x^t\in \Xc^t} \sum_{T\in \Ic_{\rm max}(\Gamma_t):x^t\in T} \kappa_T,  \label{eq:maxl:hard:distortion:optimization:objective}  \\
\text{subject to} \quad &\sum_{T\in \Ic_{\rm max}(\Gamma_t)}\kappa_T=1,  \label{eq:maxl:hard:distortion:optimization:constraint}  \\
&\kappa_T\in [0,1], \quad \forall T\in \Ic_{\rm max}(\Gamma_t).  \label{eq:maxl:hard:distortion:optimization:constraint:nonnegative}
\end{align}
\end{subequations}
\end{lemma}

On the other hand, for any $t$, $\chi_f(\Gamma_t)$ is the solution to the following linear program \cite[Section 2.2]{arbabjolfaei2018fundamentals}:  %optimization problem: 
\begin{subequations}  \label{eq:maxl:chif:optimization}
\begin{align}
\text{minimize} \quad &\sum_{T\in \Ic_{\rm max}(\Gamma_t)} \lambda_T,  \label{eq:maxl:chif:optimization:objective}  \\
\text{subject to} \quad &\sum_{T\in \Ic_{\rm max}(\Gamma_t):x^t\in T} \lambda_T\ge 1, \quad \forall x^t\in \Xc^t,  \label{eq:maxl:chif:optimization:constraint}  \\
&\lambda_T\in [0,1], \quad \forall T\in \Ic_{\rm max}(\Gamma_t).  \label{eq:maxl:chif:optimization:constraint:nonnegative}
\end{align}
\end{subequations}

We can show that the solutions to the optimization problems \eqref{eq:maxl:hard:distortion:optimization} and \eqref{eq:maxl:chif:optimization} are reciprocal to each other. That is, 
\begin{align}
\eta=\frac{1}{\chi_f(\Gamma_t)},  \label{eq:proof:thm:maxl:key}
\end{align} 
whose proof is presented in Appendix \ref{app:eq:proof:thm:maxl:key}. 
The remaining proof of Theorem \ref{thm:maxl} follows easily from the above results. 
%subsequently, %by Lemmas \ref{lem:maxl:merging:consequence} and \ref{lem:maxl:hard:distortion}, 
\begin{IEEEproof}[Proof of Theorem~\ref{thm:maxl}]
We have 
\begin{align*}
\Lc&%=\lim_{t\to \infty} \frac{1}{t} \Lc^t
\stackrel{(a)}{=}\lim_{t\to \infty} \frac{1}{t} (-\log \eta)\stackrel{(b)}{=}\lim_{t\to \infty}\frac{1}{t} \log \chi_f(\Gamma_t)\stackrel{(c)}{=}\log \chi_f(\Gamma). 
\end{align*}
where (a) follows from \eqref{eq:def:max:leakage:rate} and Lemma \ref{lem:maxl:hard:distortion}, 
(b) follows \eqref{eq:proof:thm:maxl:key}, 
and (c) follows from the fact that $\chi_f(\Gamma_t)=\chi_f(\Gamma^{\lor t})=\chi_f(\Gamma)^t$ (cf. \cite[Corollary 3.4.2]{scheinerman2011fractional}). 
\end{IEEEproof}

Having characterized the optimal maximal leakage rate $\Lc$ in Theorem \ref{thm:maxl}, 
\Royy{in the following we} design an optimal mapping scheme $P_{Y|X^t}$ for some $t$ that achieves $\Lc$, \Royy{which is based on the optimal fractional coloring of the confusion graph $\Gamma$}. 
%Towards this end, we construct a mapping scheme $P_{Y|X}$ according to the optimal fractional coloring of the confusion graph $\Gamma$. 
%We review several basic definitions regarding the coloring of a graph.
 
%\Roy{definitions for coloring, chromatic number, etc. to be added.}

%The mapping scheme is described in detail as follows. 

Fix the sequence length $t=1$. 
For $\Gamma_1=\Gamma$, there always exists some $b$-fold coloring $\Pcal=\{ T_1,T_2,\dots,T_{m} \}$ for some finite positive integer $b$ such that $\chi_f(\Gamma)=m/b$ \Rarxiv{(cf. Definitions \ref{def:b:fold:coloring} and \ref{def:fractional:chromatic}; see also \cite[Corollary 1.3.2 and Section 3.1]{scheinerman2011fractional})}. 
%(cf. Definitions \ref{def:b:fold:coloring} and \ref{def:fractional:chromatic} in Appendix \ref{app:basic:graph:theoretic:notions}; also see \cite[Section 3.1]{scheinerman2011fractional} for the proof of the existence). 

%\footnote{The existence of such a finite positive integer $b$ and a $b$-fold coloring $\Pcal=(T_1,\dots,T_{m})$ with $\chi_f(\Gamma)=m/b$ can be proved utilizing \cite[Corollary 1.3.2]{scheinerman2011fractional}. } 
%\Roy{Existence?}
%Note that every element in $\Pcal$ is a set of vertices of $\Gamma$ forming an independent set, which could be mapped to the same color. 

%For any vertex $\in \Xc$, let $c(x)$ be the set of sets within $\Pcal$ that includes $x$. In other words, $c(x)$ denotes the set of colors assigned to $x$. 

Set $\Yc=\Pcal$ (and thus every codeword $y\in \Yc$ is actually an independent set of $\Gamma$). 
Set 
\begin{align}
P_{Y|X}(y|x)=\begin{cases}1/b, \quad \text{if $x\in y$},\\0, \qquad \text{otherwise.}\end{cases}  \label{eq:maxl:mapping:scheme}
\end{align}
%As $|c(x)|=b$ for any $x\in \Xc$ (i.e., every $x\in \Xc$ is in exactly $b$ sets within $\Pcal$), we have
As every  $x\in \Xc$ is in exactly $b$ sets within $\Pcal$, we have 
\begin{align*}
\sum_{y\in \Yc}P_{Y|X}(y|x)=\sum_{y\in \Yc:x\in y}\frac{1}{b}+\sum_{y\in \Yc:x\notin y}0=1, \quad \forall x\in \Xc,
%\sum_{y\in \Yc}P_{Y|X}(y|x)=\sum_{y\in \Yc(x)}\frac{1}{b}+\sum_{y\notin \Yc(x)}0=1, \quad \forall x\in \Xc,
\end{align*}
and thus $P_{Y|X}$ is a valid mapping scheme. 
%With such mapping, we have 
We have 
\begin{align}
L^1(P_{Y|X})&=\log \sum_{y\in \Yc} \max_{x\in \Xc} P_{Y|X}(y|x)  \nonumber  \\
&=\log \sum_{y\in \Pcal} \frac{1}{b}=\log \frac{m}{b}=\log \chi_f(\Gamma),  \label{eq:maxl:mapping:scheme:efficacy}
\end{align}
and thus we know that the maximal leakage rate in Theorem~\ref{thm:maxl} is indeed achievable by the mapping described in \eqref{eq:maxl:mapping:scheme}.

\begin{remark}  \label{remark:1}
%According to \cite[Theorem 9.2]{csiszar2011information}, we know that under the worst-cast $P_X$, the optimal source coding rate is simply equal to the zero-error compression rate for the source with the confusion graph $\Gamma$. 
Consider any source coding problem $\Gamma$. 
We know that the optimal zero-error compression rate (with fixed-length deterministic source codes) is
\begin{align*}
\Rc%=\max_{P_X}H(\Gamma,P_X)
&=\lim_{t\to \infty}\frac{1}{t}\log \chi(\Gamma^{\lor t})=\lim_{t\to \infty}\frac{1}{t}\log \chi_f(\Gamma^{\lor t})=\log \chi_f(\Gamma), %\label{eq:graph:entropy:worst:case}
\end{align*}
where the second equality follows from \cite[Corollary 3.4.3]{scheinerman2011fractional}. 
%\Roy{deterministic compression is enough}
We can verify that the above result holds even when we allow stochastic mapping. 
Hence, the maximal leakage rate $\Lc$ always equals to the optimal compression rate $\Rc$. 
%To the best of our knowledge, this is the first time the notion of maximal leakage is shown to be equal to a physical measurable entity \Roy{this statement can be tricky, for example see \cite[Theorem 8]{issa2019operational}}. 
Moreover, it can be verified that any $\Rc$-achieving deterministic code can simultaneously achieve $\Lc$. 
In other words, when considering fixed-length source coding, there is no trade-off between the compression rate and the leakage rate. 
%Nevertheless, there are several things to note: 
Furthermore, we observe the following: 
\begin{enumerate}
\item Our characterization of $\Lc$ holds generally and does not rely on the assumption of fixed-length coding; 
\item While in general, the optimal zero-error compression rate $\mathcal{R}$ and the maximal leakage rate $\mathcal{L}$ can be simultaneously and asymptotically attained at the limit of increasing $t$, we showed in \eqref{eq:maxl:mapping:scheme:efficacy} that $\mathcal{L}$, on the other hand, can be achieved exactly even with $t=1$ (using the symbol-by-symbol encoding scheme specified in \eqref{eq:maxl:mapping:scheme} based on the factional coloring of $\Gamma$), but possibly at the expense of the compression rate. 
%Our optimal mapping scheme achieving $\Lc$ based on fractional coloring of $\Gamma$ is a scalar scheme where each of the $t$ symbols can be independently mapped to a codeword, while in general $\Rc=\log \chi_f(\Gamma)$ is achieved asymptotically; 
\item For variable-length source coding, whether there is a compression-leakage trade-off remains unclear. %and remains to be studied in future works.
\end{enumerate}
%\Roy{note that this is because $\log \chi(\Gamma^{\lor t})$ and $\log \chi_f(\Gamma^{\lor t})$ are equivalent when $t\to \infty$}
\end{remark}

\section{Extensions on the Maximal Leakage Rate: Multiple and Approximate Guesses}  \label{sec:extensions}

In general, the adversary may be able to make multiple guesses. For example, the adversary may possess a testing mechanism to verify whether its guess is correct or not, and thus can perform a trial and error attack until it is stopped by the system. 
Also, for each true source sequence, there may be multiple estimators other than the true sequence itself that are ``close enough" and thus can be regarded successful. 
%To provide a generalized definition of information leakage, consider the following. 
%that can be easily specialized to different adversarial models, consider the following. 

We generalize our definition of information leakage to cater to the above scenarios. 
Consider any source coding problem $\Gamma$, sequence length $t$, and valid mapping $P_{Y|X^t}$. 
Suppose the adversary generates a set of guesses $K\subseteq \Xc^t$. 
%Let \Royy{$K^+$} denote the set of true source sequences such that the at least one of the adversary's guess in $K$ can be regarded as successful. 
For each set $K$, define a ``covering" set $K^+$, where $K \subseteq K^+ \subseteq \mathcal{X}^t$, such that if the true sequence is in $K^+$, then the adversary's guess list $K$ is considered successful. 
Let %$\Sc=\{ K^+:K\subseteq \Xc^t \}$ be the collection of all possible $K^+$ formed by all possible guess lists $K$ that the adversary can choose. 
\begin{align*}
\Sc\doteq \{ K^+: \text{$K$ is a guess list the adversary can choose} \}
\end{align*}
be the collection of all possible $K^+$. 
%
%\Roy{change $S\in \Sc$ to $K^+\in \Sc$.}
Then for the blind guessing, the successful probability is $\max_{S\in \Sc} \sum_{x^t\in S} P_{X^t}(x^t),$ and for guessing after observing $Y$, the average successful probability is ${\mathbb E}_Y \left[ \max_{S\in \Sc} \sum_{x^t\in S} P_{X^t|Y}(x^t|Y) \right] .$
In the same spirit of maximal leakage, we can define 
\begin{align*}
\rho_t(P_{Y|X^t},\Sc)%  \nonumber  \\
&\doteq \log \max_{P_X} \frac{ {\mathbb E}_Y \hspace{-0mm}\left[ \max\limits_{S\in \Sc} \sum\limits_{x^t\in S} P_{X^t|Y}(x^t|Y) \right] }{\max\limits_{S\in \Sc} \sum\limits_{x^t\in S} P_{X^t}(x^t)}  %\label{eq:def:unified:leakage}
\end{align*}
as the ratio between the a posteriori and a priori successful guessing probability. 
%the expected successful guessing probability after and before observing $Y$. 
If we set \Royy{$\Sc_{\rm singleton}=\{ \{x^t\}: x^t\in \Xc^t \}$}, 
that is, the adversary is allowed one guess and it must guess the correct source sequence precisely, 
the maximal leakage defined in \eqref{eq:def:maxl:leakage:rate:any:t:mapping:operational} can be equivalently written as 
\begin{align*}
L^t(P_{Y|X^t})&=\rho_t(P_{Y|X^t},\Sc_{\rm singleton}).
\end{align*}

%\Royy{In the next two subsections, we study different scenarios where the adversary makes multiple guesses allowing no distortion and one guess allowing certain distortion, respectively}. 
%\Royy{For the investigation of the leakage under the generic adversarial model where the adversary makes multiple guesses allowing distortion, see the full version of this paper \cite{isit:2021:arxiv}}. 

\Rarxiv{In the next three subsections, we study the information leakage rate under different adversarial models.}

\subsection{Leakage for the Case of Multiple Guesses}  \label{sec:multiple:precise:guesses}

%In this subsection, we consider the case when the adversary is able to make multiple guesses after each observation of the message. 
In this subsection, we consider the case where the adversary make multiple guesses, yet does not allow distortion between its estimators and the true sequence. 
\iffalse
In general, the adversary may be able to make multiple guesses after each observation of the codeword. 
For example, the adversary may possess a testing mechanism to verify whether its estimator is correct or not, and thus can perform a trial and error attack until it is stopped by the system. 
%\Roy{The adversary may have some test mechanism; cite some papers about guessing or brute-force adversary.}
\fi

%More specifically, 
%consider any function $g(t)$ characterising the number of guesses the adversary can make, % given sequence length $t$, 
We characterize the number of guesses the adversary can make by a \emph{guessing capability} function $g(t)$, 
where $t\in {\mathbb Z}^+$ is the sequence length. 
We assume $g(t)$ to be positive, integer-valued, non-decreasing, and upper-bounded\footnote{Suppose for some $t$ we have $g(t)\ge \alpha(\Gamma_t)$. Then upon observing any codeword $y$, the adversary can always determine the true source value by exhaustively guessing all possible $x^t\in \Xc^t(y)$ as $|\Xc^t(y)|\le \alpha(\Gamma_t)$.} by $\alpha(\Gamma_t)=\alpha(\Gamma^{\lor t})=\alpha(\Gamma)^t$, 
where $\alpha(\cdot)$ denotes the independence number of a graph \Rarxiv{(cf. Appendix \ref{app:basic:graph:theoretic:notions})}. 
%for every $t\le t'$ we have $k(t),k(t')\in {\mathbb Z}^+$, $1\le k(t)\le k(t')\le \alpha(\Gamma_t)$. 
%The information leakage will now be characterized by the ratio between the sum probability of the $g(t)$ most likely source sequences after and before observing the transmitted codeword. 

%For a given $P_X$, we define the $k$-guess leakage rate as 
%\begin{align}
%\Lc^{k,*}[P_X]=\lim_{\e\to 0}\lim_{t\to \infty}\inf_{P_{Y|X^t}:P_e\le \e}\frac{1}{t}\log \frac{\sum_y\max_{K\subseteq \Xc^t:|K|\le k(t)}\sum_{x\in K}P_{X^t,Y}(x,y)}{\max_{K\subseteq \Xc^t:|K|\le k(t)}P_{X^t}(K)},
%\end{align}
%and the worst-cast $k$-guess leakage rate as

Consider any source coding problem $\Gamma$ and any guessing-capability function $g$. 
For a given sequence length $t$ and a given valid mapping $P_{Y|X^t}$, the maximal leakage naturally extends to the \emph{multi-guess} maximal leakage, defined as 
%We define the \emph{multi-guess} maximal leakage for a given sequence length $t$ and a given valid mapping scheme $P_{Y|X^t}$, denoted by $\Lc(\Gamma,g,t,P_{Y|X^t})$, as 
\begin{align}
L^t_g(P_{Y|X^t})\doteq \rho_t(P_{Y|X^t},\Sc_g),  \label{eq:def:multiple:guesses:any:t:mapping}
\end{align}
where $ \Sc_{g}=\{ K\subseteq \Xc^t: |K|=g(t) \}$. Then we can define the (optimal) multi-guess maximal leakage rate as 
\begin{align}
\Lc_g&\doteq \lim_{t\to \infty} \frac{1}{t} \inf_{\substack{P_{Y|X^t}: \\ \Royy{\Xc^t(y)\in \Ic(\Gamma_t)},\forall y\in \Yc}}L^t_g(P_{Y|X^t}).  \label{eq:def:multuple:guesses:rate}
\end{align}

We first show that the multi-guess maximal leakage rate is always no larger than the maximal leakage. 
%In fact, as we shall show by the following lemma, while the absolute value of posteori successful guessing possibility becomes larger when the adversary can make more than one guess, the multiplicative gain actually 

\begin{lemma}  \label{lem:multiple:guesses:less:than:maxl}
For any source coding problem $\Gamma$ and guessing capability function $g$, we have 
$\Lc_g\le \Lc$. 
\end{lemma}
\begin{IEEEproof}
It suffices to show $L_g^t(P_{Y|X^t})\le L^t(P_{Y|X^t})$ for any $t$ and $P_{Y|X^t}$. For any $P_X$, we have 
\begin{align}
&\frac{\sum\limits_{y\in \Yc}\max\limits_{K\subseteq \Xc^t:|K|=g(t)}\sum\limits_{x^t\in K}P_{X^t,Y}(x^t,y)}{\max\limits_{\substack{K\subseteq \Xc^t:|K|=g(t)}} \sum_{x^t\in K} P_{X^t}(x^t)}  \nonumber  \\
%%%
%&=\frac{\sum\limits_{y\in \Yc}\max\limits_{K\subseteq \Xc^t:|K|\le g(t)}\sum\limits_{x^t\in K}P_{X^t}(x^t)P_{Y|X^t}(y|x^t)}{\max\limits_{\substack{K\subseteq \Xc^t:|K|\le g(t)}} \sum_{x^t\in K} P_{X^t}(x^t)}  \nonumber  \\
%%%
&\le \frac{\sum\limits_{y\in \Yc} \big( \max\limits_{K\subseteq \Xc^t:|K|=g(t)}\sum\limits_{x^t\in K}P_{X^t}(x^t) \big) \big( \max\limits_{\xt^t\in \Xc^t}P_{Y|X^t}(y|\xt^t) \big)}{\max\limits_{\substack{K\subseteq \Xc^t:|K|=g(t)}} \sum_{x^t\in K} P_{X^t}(x^t)}  \nonumber  \\
&=\sum_{y\in \Yc}\max_{x^t\in \Xc^t}P_{Y|X^t}(y|x^t),  \nonumber
\end{align}
which implies that $L_g^t(P_{Y|X^t})\le L^t(P_{Y|X^t})$. %and thus completes the proof. 
\end{IEEEproof}

We have the following single-letter lower and upper bounds on $\Lc_g$, 
%whose proof can be found in \cite{isit:2021:arxiv}. 
whose proof is presented in Appendix \ref{app:proof:thm:multiple:guesses}. 
\begin{theorem}  \label{thm:multiple:guesses}
We have
\begin{align}
\log |V(\Gamma)|-\log \alpha(\Gamma)\le \Lc_g \le \log \chi_f(\Gamma). 
\end{align}
\end{theorem}

When the adversary guesses some randomized function $U$ of $X$ rather than $X$ itself, the maximal leakage equals to its multi-guess extension \cite{issa2019operational}. 
It remains to be investigated whether a similar equivalence holds generally in our setup. 
%\end{remark}
%While it is unclear whether $\Lc_g=\Lc$ for a general source coding problem $\Gamma$ and a general guessing-capability function $g$, 
In the following we recognize one special case where \Royy{indeed} $\Lc_g=\Lc$ and consequently, by Theorem \ref{thm:maxl}, $\Lc_g=\log \chi_f(\Gamma)$. 
%the lower and upper bounds in Theorem \ref{thm:multiple:guesses} match and thus lead to 

%\begin{remark}  \label{rmk:limit:of:g}
%Set $\sigma\doteq \lim_{t\to \infty}\frac{1}{t}\log g(t)$. When $\sigma=0$, we actually have $\Lc(\Gamma,g)=\Lc(\Gamma)$. 
%\end{remark}

\begin{proposition}  \label{prop:multiple:guesses:equal:to:maxl:small:g}
\Royy{Consider any source coding problem $\Gamma$.
%For any guessing capability function $g$ such that 
If $\lim_{t\to \infty}\frac{1}{t}\log g(t)=0$, then
$\Lc_g=\Lc=\log \chi_f(\Gamma)$}. 
\end{proposition}

The proof of the above proposition 
%is presented in \cite{isit:2021:arxiv}. 
is relegated to Appendix \ref{app:proof:prop:multiple:guesses:equal:to:maxl:small:g}. 
Intuitively, the above result suggests that when the number of guesses the adversary can make does not grow ``fast enough" with respect to $t$, it makes no difference whether the adversary is making one guess or multiple guesses (in terms of the leakage defined in \eqref{eq:def:max:leakage:rate} and \eqref{eq:def:multuple:guesses:rate}). 

As a direct corollary of Theorem \ref{thm:multiple:guesses}, the result below shows that $\Lc_g=\Lc=\log \chi_f(\Gamma)$ holds for another specific scenario. 

\begin{corollary}
%Consider any zero-error source coding problem whose confusion graph $\Gamma$ is vertex-transitive\footnote{While the definition in \cite[Section 1.3]{scheinerman2011fractional} is for a hypergraph, it can be readily specialized to a graph since any graph is a special hypergraph, whose every hyperedge is a $2$-element set. } \cite[Section 1.3]{scheinerman2011fractional} and any guessing capability function $g$. 
\Royy{If $\Gamma$ is vertex-transitive\footnote{While the definition in \cite[Section 1.3]{scheinerman2011fractional} is for a \emph{hypergraph}, it can be readily specialized to a graph since any graph is a special hypergraph, whose every hyperedge is a $2$-element set. } \cite[Section 1.3]{scheinerman2011fractional}, then $\Lc_g=\Lc=\log \chi_f(\Gamma)$ for any function $g$}. 
\end{corollary}

\begin{IEEEproof}
Since $\Gamma$ is vertex-transitive, by \cite[Proposition 3.1.1]{scheinerman2011fractional}, we have $ \chi_f(\Gamma)=|V(\Gamma)|/\alpha(\Gamma)$, which indicates that lower and upper bounds in Theorem \ref{thm:multiple:guesses} match with each other, thus establishing $\Lc_g=\log \chi_f(\Gamma)=\Lc$. 
\iffalse
\begin{align*}
\log \chi_f(\Gamma)=\log \frac{|\Xc|}{\alpha(\Gamma)}=\log |\Xc|-\log \alpha(\Gamma).
\end{align*}
That is, the lower and upper bounds in Theorem \ref{thm:multiple:guesses} match, and thus establish that $\Lc_g=\log \chi_f(\Gamma)=\Lc$. 
\fi
\end{IEEEproof}

%\vspace{5mm}

\subsection{Leakage for the Case of One Approximate Guess}  \label{sec:one:approximate:guess}

%In this subsection, we consider an extension to the maximal leakage in another direction. 

Suppose that the adversary makes only one guess yet 
allows a certain level of distortion between its estimator and the true source value. 
That is, the guess is regarded successful as long as the estimator is an acceptable \emph{approximation} to the true value. 
Inspired by the notion of confusion graph $\Gamma$ that characterizes the distinguishability within the source symbols,  
we introduce another graph to characterize the approximation relationship among source symbols (from the adversary's perspective). 
%
%only needs the guess to be close enough to the true source value. %within certain distance from the true source value. 
%Similar to the confusion graph $\Gamma$ defined on $V(\Gamma)=\Xc$ which characterises the confusability between any two source values, 
%the approximation relationships (from the adversary's perspective) among the source symbols can be charaterized by an \emph{(adversary's) approximation graph} $\Theta$. 
We call this graph the adversary's approximation graph, or simply the approximation graph, denoted by $\Theta$. 
The vertex set of $\Theta$ is just the source alphabet, i.e., $V(\Theta)=\Xc$, and any two source symbols $x\neq x'\in \Xc$ are acceptable approximations to each other iff they are adjacent in $\Theta$, i.e., $\{x,x'\}\in \Ec(\Theta)$. 

Given a sequence length $t$, any two sequences $x^t=(x_1,\ldots,x_t)$ and $v^t=(v_1,\dots,v_t)$ are acceptable approximations to each other iff for every $j\in [t]$, $x_j=v_j$ or $\{x_j,v_j\}\in \Ec(\Theta)$. 
Hence the approximation graph $\Theta_t$ for sequence length $t$ is the $t$-th power of $\Theta$ with respect to the \emph{AND graph product} \cite[Section 5.2]{hammack2011handbook}: $\Theta_t={\Theta \boxtimes \Theta \boxtimes \cdots \boxtimes \Theta}=\Theta^{\boxtimes t}$. 

%\begin{definition}[AND product, \cite{scheinerman2011fractional}]  \label{def:and:product}
%Given two undirected graphs $G_1,G_2$, the AND product $G=G_1\boxtimes G_2$ is constructed as $V(G)=V(G_1)\times V(G_2)$ and $\{ (i_1,i_2),(j_1,j_2) \}\in \Ec(G)$ iff the following two conditions are simultaneously satisfied:
%\begin{enumerate} 
%\item $\{ i_1,j_1 \}\in \Ec(G_1)$ or $i_1=i_2$. 
%\item $\{ j_1,j_2 \}\in \Ec(G_2)$ or $j_1=j_2$.
%\end{enumerate}
%\end{definition}

For any vertex $x^t\in \Xc^t$, let $N(\Theta_t,x^t)$ denote the \emph{neighborhood} of $x^t$ within $\Theta_t$, including the vertex $x^t$ itself. That is, $N(\Theta_t,x^t)=\{ v^t\in \Xc^t:\text{$v^t=x^t$ or $\{v^t,x^t\}\in \Ec(\Theta_t)$} \}$.\footnote{\Royy{This is referred to as the \emph{closed} neighborhood of $x^t$ in $\Theta_{x^t}$ in \cite{scheinerman2011fractional}, in contrast to the \emph{open} neighborhood of $x^t$ which does not include $x^t$ itself.}}

%For a given $P_X$, we define the approximate-guess leakage rate as 
%\begin{align}
%\Lc^{*,\Theta}[P_X]=\lim_{\e\to 0}\lim_{t\to \infty}\inf_{P_{Y|X^t}:P_e\le \e}\frac{1}{t}\log \frac{\sum_y \max_{x\in \Xc^t}P_{X^t,Y}(N(\Theta_t,x),y)}{\max_{x\in \Xc^t}P_{X^t}(N(\Theta,x))},
%\end{align}
%where $N(\Theta_t,x)$ denotes the neighbours of $x$ in $\Theta_t$ (including $x$ itself). 

Consider any source coding problem $\Gamma$ and any approximation graph $\Theta$. 
For a given sequence length $t$ and a given valid mapping $P_{Y|X^t}$, the maximal leakage naturally extends to the \emph{approximate-guess} maximal leakage, defined as 
\begin{align}
&L^t_{\Theta}(P_{Y|X^t})\doteq \rho_t(P_{Y|X^t},\Sc_{\Theta}),  \label{eq:def:one:approximate:guess:any:t:mapping}
\end{align}
where $ \Sc_{\Theta}=\{ N(\Theta_t,x^t): x^t\in \Xc^t \}. $
Then we can define the (optimal) approximate-guess maximal leakage rate as
\begin{align}
\Lc_{\Theta}
&\doteq \lim_{t\to \infty} \frac{1}{t} \inf_{\substack{P_{Y|X^t}: \\ \Royy{\Xc^t(y)\in \Ic(\Gamma_t)},\forall y\in \Yc}} L^t_{\Theta}(P_{Y|X^t}). 
\end{align}

The approximate-guess maximal leakage is always no larger than the maximal leakage as specified in lemma below, whose proof is similar to that of Lemma \ref{lem:multiple:guesses:less:than:maxl} and thus omitted.

\begin{lemma}  \label{lem:one:approximate:guess:less:than:maxl}
For any source coding problem $\Gamma$ and approximation graph $\Theta$, we have $\Lc_{\Theta}\le \Lc$. 
\end{lemma}

%\begin{IEEEproof}
%It suffices to show $\Lc(\Gamma,\Theta,t,P_{Y|X^t})\le \Lc(\Gamma,t,P_{Y|X^t})$ for any $t$ and $P_{Y|X^t}$. For any $P_X$, we have 
%\begin{align}
%&\frac{\sum\limits_{y\in \Yc} \max\limits_{x^t\in \Xc^t} \sum\limits_{\xt^t\in N(\Theta_t,x^t)} P_{X^t,Y}(\xt^t,y)}{\max\limits_{x^t\in \Xc^t} \sum\limits_{\xt^t\in N(\Theta_t,x^t)} P_{X^t}(\xt^t)}  \nonumber  \\
%%%
%%&=\frac{\sum\limits_{y\in \Yc} \max\limits_{x^t\in \Xc^t} \sum\limits_{\xt^t\in N(\Theta_t,x^t)} P_{X^t}(\xt^t)\cdot P_{Y|X^t}(y|\xt^t)}{\max\limits_{x^t\in \Xc^t}P_{X^t}(N(\Theta_t,x^t))}  \nonumber  \\
%%%%
%&\le \frac{\sum\limits_{y\in \Yc} \big( \max\limits_{x^t\in \Xc^t} \sum\limits_{\xt^t\in N(\Theta_t,x^t)} P_{X^t}(\xt^t) \big) \big( \max\limits_{v^t\in \Xc^t} P_{Y|X^t}(y|v^t) \big)}{\max\limits_{x^t\in \Xc^t} \sum\limits_{\xt^t\in N(\Theta_t,x^t)} P_{X^t}(\xt^t)}  \nonumber  \\
%%%
%&=\sum_{y\in \Yc} \max_{x^t\in \Xc^t} P_{Y|X^t}(y|x^t),  \nonumber
%\end{align}
%which indicates that $\Lc(\Gamma,\Theta,t,P_{Y|X^t})\le \Lc(\Gamma,t,P_{Y|X^t})$ and thus completes the proof. 
%\end{IEEEproof}

%In the following we present single-letter lower and upper bounds on $\Lc^*(\Gamma,\Theta)$. Towards that end, we introduce the following graph-theoretic notions. 
Before presenting single-letter bounds on $\Lc_{\Theta}$, we introduce the following graph-theoretic notions. 

Consider any source coding problem $\Gamma$, approximation graph $\Theta$, and sequence length $t$. 
For any maximal independent set $T\in \Ic_{\rm max}(\Gamma_t)$, we define its \emph{associated hypergraph} (\Rarxiv{see Appendix \ref{app:basic:graph:theoretic:notions} for basic definitions about hypergraphs}). 
%(\Royy{see \cite[Chapter 1]{scheinerman2011fractional} for basic definitions about hypregraphs}). 

\iffalse
\begin{definition}[Hypergraph, \cite{scheinerman2011fractional}]  \label{def:hypergraph}
\Royy{A hypergraph $\Hc$ consists of a vertex set $V(\Hc)$ and a hyperedge set $\Ec(\Hc)$, which is a family of subsets of $V(\Hc)$}. 
%A hypergraph $\Hc$ is a pair $(V,\Ec)$, where $V$ is a finite set, called the vertex set of $\Hc$, and $\Ec$ is a family of subsets of $V$, called the hyperedge set of $\Hc$. 
%The sets $V$ and $\Ec$ may also be denoted by $V(\Hc)$ and $\Ec(\Hc)$, respectively. 
\end{definition}
\fi

%For more details on hypergraph, see \cite[Chapter 1]{scheinerman2011fractional}. %Appendix \ref{app:basic:graph:theoretic:notions}.  

\begin{definition}[Associated Hypergraph]  \label{def:associated:hypergraph}
%For any source coding problem $\Gamma$, approximation graph $\Theta$, sequence length $t$, and maximal independent set $T\in \Ic_{\rm max}(\Gamma_t)$, define the associated hypergraph $\Hc_T$ as 
Consider any sequence length $t$. 
For any $T\in \Ic_{\rm max}(\Gamma_t)$, its associated hypergraph\footnote{Note that, for brevity, the dependence of the associate hypergraph on the underlying approximation graph $\Theta$ is not shown in the notation $\Hc_t(T)$.} $\Hc_t(T)$ is defined as $V(\Hc_t(T))=T$ and $\Ec(\Hc_t(T))=\{ E\subseteq T: E\neq \emptyset, \text{$E=T\cap N(\Theta_t,x^t)$ for some $x^t\in \Xc^t$} \}$. 
%\begin{itemize}
%\item $V(\Hc_t(T))=T$.
%\item $\Ec(\Hc_t(T))=\{ T\cap N(\Theta_t,x^t):x^t\in \Xc^t \}$.
%\end{itemize}
\end{definition}

%Note that the above hyperedge set is a \emph{multiset} \cite{blizard1988multiset} and may include empty sets. 

%Note that the above hyperedge set may include the empty set.
 
%Based on the above definition, 
%we have the following single-letter lower and upper bounds on $\Lc^*(\Gamma,\Theta)$. 

%\begin{theorem}
%We have
%\begin{align}
%&\log \frac{1}{(\max_{T\in \Ic_{\rm max}(\Gamma)}K_f(T))(\max_{x\in X}P_X{N(\Theta,x)})}  \nonumber  \\
%&\qquad \le \Lc^{*,\Theta}[P_X] \le \log \chi_f(\Gamma)\frac{\max_{x\in Xc,T\in \Ic_{\rm max}(\Gamma)}P_X(N(\Theta,x)\cap T)}{\max_{x\in \Xc} P_X(N(\Theta,x))}
%\end{align}
%\end{theorem}

The following \Royy{single-letter} lower and upper bounds on $\Lc_{\Theta}$ hold, 
%whose proof can be found in \cite{isit:2021:arxiv}. 
whose proof is presented in Appendix \ref{app:proof:thm:one:approximate:guess}. 

\begin{theorem}  \label{thm:one:approximate:guess}
We have
\begin{align}
%&\log \frac{1}{\max\limits_{T\in \Ic_{\rm max}(\Gamma)}k_f(\Hc_1(T))} + \log \max_{P_X} \min_{x\in \Xc} \frac{1}{P_X(N(\Theta,x))}  \nonumber  \\
%&\log p_f(\Theta) - \log \max\limits_{T\in \Ic_{\rm max}(\Gamma)}k_f(\Hc_1(T))  \nonumber  \\
&\log \frac{p_f(\Theta)}{\max\limits_{T\in \Ic_{\rm max}(\Gamma)}k_f(\Hc_1(T))}\le \Lc_{\Theta} \le  \log \chi_f(\Gamma), 
\end{align}
where $p_f(\cdot)$ denotes the fractional closed neighborhood packing number \cite[Section 7.4]{scheinerman2011fractional} of a graph and $k_f(\cdot)$ denotes the fractional covering number \Rarxiv{(cf. Definition \ref{def:fractional:covering:number}) of a hypergraph}.
%\cite[Section 1.2]{scheinerman2011fractional} of a hypergraph. 

%\begin{align}
%k_f(\Gamma,\Theta)=\max_{T\in \Ic_{\rm max}(\Gamma)}k_f(\Hc_T). \label{eq:def:kf:gamma:theta}
%\end{align} 
%and for any hypergraph $\Hc$, $k_f(\Hc)$ denotes its fractional covering number. 
\end{theorem}

\begin{remark}
While the lower bound in Theorem \ref{thm:one:approximate:guess} takes both $\Gamma$ and $\Theta$ into account, the upper bound solely depends on $\Gamma$. 
\end{remark}

%\begin{remark}
%We have also studied the leakage under the generic adversarial model where the adversary makes multiple approximate guesses. See the full version of this paper \cite{isit:2021:arxiv}. 
%\end{remark}

%Having analyzed $\Lc_g$ and $\Lc_{\Theta}$, it is only natural to move on to the analysis of the leakage under the generic adversarial model where the adversary makes multiple approximate guesses

%\vspace{5mm}

%\begin{remark}
%\Roy{Leakage for the Case of Multiple Approximate Guesses}
%\end{remark}

%%%%%%%%%%%%%%%%%%%%%%%%%%%%%%%%%%%%
\subsection{Leakage for the Case of Multiple Approximate Guesses}  \label{sec:multiple:approximate:guesses}

%\Roy{this subsection to be removed from the ISIT version}

We consider the most generic mathematical model so far by allowing the adversary to make multiple guesses after each observation of the codeword, and a guess is regarded as successful as long as the estimated sequence is in the neighborhood of the true source sequence. 

Consider any source coding problem $\Gamma$, approximation graph $\Theta$, and guessing capability function $g$. 
Note that we require function $g$ to be upper bounded\footnote{Suppose for some $t$ we have $g(t)\ge \max_{T\in \Ic_{\rm max}(\Gamma_t)}k(\Hc_t(T))$. Upon observing any $y$, there exists some covering of $\Xc^t(y)$ with no more than $ \max_{T\in \Ic_{\rm max}(\Gamma_t)}k(\Hc_t(T))$ hyperedges, each corresponding to one unique vertex $x^t$ (cf. Definition \ref{def:associated:hypergraph}). The adversary can simply choose these $x^t$ as its estimator and the probability of successfully guessing will be $1$.}  
%The reason is similar to the one stated in the multi-guess scenario in Section \ref{sec:multiple:precise:guesses}. 
%always determine the true source value by exhaustively guessing all possible $x^t\in \Xc^t(y)$ as $|\Xc^t(y)|\le \alpha(\Gamma_t)$.
as $$g(t)\le \max_{T\in \Ic_{\rm max}(\Gamma_t)}k(\Hc_t(T)), \quad \forall t\in {\mathbb Z}^+, $$
where $k(\cdot)$ denotes the covering number \Rarxiv{(cf. Definition \ref{def:covering})} of a hypergraph.
%of a hypergraph (cf. \cite[Section 1.1]{scheinerman2011fractional}). 

For any sequence length $t$ and valid mapping $P_{Y|X^t}$, 
the\emph{ multi-approximate-guess} maximal leakage is defined as
%Define shorthand notation $N(\Theta_t,K) \triangleq \cup_{x^t\in K}N(\Theta_t,x^t)$. 

%\begin{align}
%\log \frac{1}{\max_{x\in \Xc}P_X(N(\Theta,x))}-\sigma+\lim_{t\to \infty} t^{-1}\log (1-(1-(\frac{1}{k_f(\Gamma,\Theta)})^t)^{k(t)})  \\
%\end{align}

\begin{align}
L^t_{\Theta,g}(P_{Y|X^t})\doteq \rho_t(P_{Y|X^t},\Sc_{\Theta,g}),
\end{align}
where $ \Sc_{\Theta,g}=\{ \cup_{x^t\in K} N(\Theta_t,x^t): K\subseteq \Xc^t, |K|=g(t) \}. $
Then we can define the (optimal) multi-approximate-guess maximal leakage rate as 
\begin{align}
&\Lc_{\Theta,g}\doteq \lim_{t\to \infty} \frac{1}{t} \inf_{\substack{P_{Y|X^t}: \\ \Xc^t(y)\in \Ic(\Gamma_t),\forall y\in \Yc}} L^t_{\Theta,g}(P_{Y|X^t}).
\end{align}

\iffalse
\begin{align}
&\Lc(\Gamma,\Theta,g,t,P_{Y|X^t})  \nonumber  \\
%%
&=\log \max_{P_X} \frac{{\mathbb E}_Y [ \max\limits_{\substack{K\subseteq \Xc^t:\\|K|\le g(t)}} \sum\limits_{x^t\in N(\Theta_t,K)} P_{X^t|Y}(x^t|Y) ] }{\max\limits_{\substack{K\subseteq \Xc^t:|K|\le g(t)}} \sum\limits_{x^t\in N(\Theta_t,K)} P_{X^t}(x^t)}  \\
%%
&=\log \max_{P_X} \frac{\sum\limits_{y\in \Yc} \max\limits_{\substack{K\subseteq \Xc^t:\\|K|\le g(t)}} \sum\limits_{x^t\in N(\Theta_t,K)} P_{X^t,Y}(x^t,y)}{\max\limits_{\substack{K\subseteq \Xc^t:|K|\le g(t)}} \sum\limits_{x^t\in N(\Theta_t,K)} P_{X^t}(x^t)},
\end{align}
where $N(\Theta_t,K)=\cup_{x^t\in K}N(\Theta_t,x^t)$. 

Based on the above definition, we define the (optimal) multi-approximate-guess maximal leakage rate as 
\begin{align}
&\Lc^*(\Gamma,\Theta,g)  \nonumber  \\
&=\lim_{t\to \infty} \frac{1}{t} \inf_{\substack{P_{Y|X^t}: \\ \Xc^t(y)\in \Ic(\Gamma_t),\forall y\in \Yc}} \Lc(\Gamma,\Theta,g,t,P_{Y|X^t}).
\end{align}
%Then we can define the optimal multi-approximate-guess maximal leakage for a given sequence length $t$ as 
%\begin{align}
%\Lc^*(\Gamma,\Theta,g,t)=\inf_{\substack{P_{Y|X^t}: \\ \Xc^t(y)\in \Ic(\Gamma_t),\forall y\in \Yc}} \Lc(\Gamma,\Theta,g,t,P_{Y|X^t}), 
%\end{align}
%and the (optimal) multi-approximate-guess maximal leakage rate as
%\begin{align}
%\Lc^*(\Gamma,\Theta,g)=\lim_{t\to \infty} \frac{1}{t} \Lc^*(\Gamma,\Theta,g,t). 
%\end{align}
\fi

%The multi-approximate-guess maximal leakage rate is always no larger than the maximal leakage rate. 
Once again, following a similar proof to Lemma \ref{lem:multiple:guesses:less:than:maxl}, we can show the result below. 
%that the multi-approximate-guess maximal leakage rate is always no larger than the maximal leakage rate. 

%One can define $\Lc(\Gamma,\Theta,k)$ as the (optimal) $k$-approximate-guess maximal leakage rate under such assumptions with given confusion graph $\Gamma$, approximation graph $\Theta$, and guessing capability function $g$. 

\begin{lemma}  \label{lem:multiple:approximate:guesses:less:than:maxl}
%For any source coding problem with confusion graph $\Gamma$, any adversary's approximation graph $\Theta$, and guessing capability function $g$, we have $\Lc(\Gamma,\Theta,g)\le \Lc(\Gamma)$. 
For any source coding problem $\Gamma$, approximation graph $\Theta$, and guessing capability function $g$, we have $\Lc_{\Theta,g}\le \Lc$. 
\end{lemma}

We have the following lower and upper bounds. % on $\Lc^*(\Gamma,\Theta,g)$.  

\begin{theorem}  \label{thm:multiple:approximate:guesses}
We have 
\begin{align}
%&\log \frac{1}{\max\limits_{T\in \Ic_{\rm max}(\Gamma)}k_f(\Hc_1(T))} + \log \max_{P_X} \min_{x\in \Xc} \frac{1}{P_X(N(\Theta,x))}  \nonumber  \\
%&\qquad \qquad \qquad \qquad \qquad \le \Lc_{\Theta,g} \le  \log \chi_f(\Gamma). 
&\log \frac{p_f(\Theta)}{\max\limits_{T\in \Ic_{\rm max}(\Gamma)}k_f(\Hc_1(T))}\le L_{\Theta,g} \le  \log \chi_f(\Gamma), 
\end{align}
%where $\sigma=\lim_{t\to \infty}t^{-1}\log k(t)$. 
\end{theorem}

The proof of the above theorem is given in Appendix \ref{app:proof:thm:multiple:approximate:guesses}.

%\begin{remark}
An interesting observation is that the lower and upper bounds for $\Lc_{\Theta,g}$ are exactly the same as the lower and upper bounds for $\Lc_{\Theta}$, respectively. 
However, we do not know whether $\Lc_{\Theta,g}=\Lc_{\Theta}$ holds in general. 
%\end{remark}
%\begin{remark}
%Recall that $\sigma\doteq \lim_{t\to \infty}\frac{1}{t}\log g(t)$. When $\sigma=0$, we actually have $\Lc(\Gamma,\Theta,g)=\Lc(\Gamma,\Theta)$. 
%\end{remark}
%Similar to Proposition \ref{d}, we have the following result. 

As specified in Proposition \ref{prop:multiple:guesses:equal:to:maxl:small:g}, when the number of guesses the adversary can make does not grow ``fast enough" with respect to $t$, or, more precisely, when $\lim_{t\to \infty}\frac{1}{t}\log g(t)=0$, the multi-guess maximal leakage rate indeed equals to the maximal leakage rate. 
The following lemma states that a similar equivalence holds even when the adversary allows approximation guesses. 

\begin{proposition}  \label{prop:multiple:approximate:guesses:equal:to:one:approximate:guess:small:g}
Consider any source coding problem $\Gamma$ and any approximation graph $\Theta$. 
For any guessing capability function $g$ such that 
%\begin{align}
$\lim_{t\to \infty}\frac{1}{t}\log g(t)=0$,  
%\end{align}
we have $\Lc_{\Theta,g}=\Lc_{\Theta}$. 
\end{proposition}

The proof is similar to that of Proposition \ref{prop:multiple:guesses:equal:to:maxl:small:g} and omitted.

%% file: proofs_isit_2021.tex
\appendices

\section{Basic Graph-Theoretic Notions}  \label{app:basic:graph:theoretic:notions}

Consider a directed, finite, simple, and undirected graph $G=(V,\Ec)$, where $V=V(G)$ is the set of vertices of $G$ and $\Ec=\Ec(G)$ is the set of edges in $G$, which is a set of $2$-element subsets of $V$. 
A edge $\{v_1,v_2\}\in \Ec(G)$ means that vertices $v_1$ and $v_2$ are adjacent in the graph $G$. 

An independent set of $G$ is a subset of vertices $T\subseteq V$ with no edge among them. 
An independent set $T$ is said to be maximal iff there exists no other independent set in $G$ that is a superset of $T$. 
For the graph $G$, let $\Ic(G)$ denote the collection of its independent sets, and let $\Ic_{\rm max}(G)=\{ T\in \Ic_{\rm max}(G):T\not \subseteq T',\forall T'\in \Ic_{\rm max}(G)\setminus \{T\} \}$ denote the collection of its maximal independent sets. 
Also, let $\alpha(G)$ denote the independence number of $G$, i.e., the size of the largest independent set in $G$. 

%\Roy{multiset definition}

We review the following basic definitions. 

A \emph{multiset} is a collection of elements in which each element may occur more than once \cite{blizard1988multiset}. The number of times an element occurs in a multiset is called its \emph{multiplicity}. For example, $\{a,a,b\}$ is a multiset, where the elements $a$ and $b$ have multiplicities of $2$ and $1$, respectively. \Rarxiv{The cardinality of a multiset is the summation of the multiplicities of all its elements}. 

\begin{definition}[Coloring and chromatic number, \cite{scheinerman2011fractional}]  \label{def:coloring}
Given a graph $G=(V,\Ec)$, a coloring of $G$ is a partition of the vertex set $V$, $\Pcal=\{T_1,T_2,\dots,T_m\}$, such that for every $j\in [m]$, $T_j\in \Ic(G)$. 
%$T_j$ is an independent set of $G$, i.e., $T_j\in \Ic(G)$. 
The chromatic number of $G$, denoted by $\chi(G)$, is the smallest integer $m$ such that a coloring $\Pcal=\{T_1,T_2,\dots,T_m\}$ exists for $G$. 
\end{definition}

\begin{definition}[$b$-fold coloring and $b$-fold chromatic number, \cite{scheinerman2011fractional}]  \label{def:b:fold:coloring}
Given a graph $G=(V,\Ec)$, a $b$-fold coloring of $G$ for some positive integer $b$ is a multiset $\Pcal=\{T_1,T_2,\dots,T_m\}$ such that for every $j\in [m]$, $T_j\in \Ic(G)$, and every vertex $v\in V$ is in exactly $b$ sets in $\Pcal$. 
The $b$-fold chromatic number of $G$, denoted by $\chi_b(G)$, is the smallest integer $m$ such that a $b$-fold coloring $\Pcal=\{T_1,T_2,\dots,T_m\}$ exists for $G$. 
\end{definition}

\begin{definition}[Fractional chromatic number, \cite{scheinerman2011fractional}]  \label{def:fractional:chromatic}
Given a graph $G=(V,\Ec)$, the fractional chromatic number $\chi_f(G)$ is defined as 
\begin{align*}
\chi_{f}(G)=\inf_{b}\frac{\chi_b(G)}{b}=\lim_{b\to \infty}\frac{\chi_b(G)}{b},
\end{align*}
where the second equality follows from the subadditivity of $\chi_b(G)$ in $b$ and Fekete's Lemma \cite{fekete1923lemma}. 
\end{definition}

%\Rarxiv{packing; packing number; fractional packing number}

%\Roy{Notation comparison between $\chi_f$ and $\chi_{\rm f}$: Since that $\chi_f$ has been commonly used in the graph theory literature including the book \cite{scheinerman2011fractional}, and that $\chi_{\rm f}$ does not look good, I decided to keep using $\chi_f$ instead of $\chi_{\rm f}$, even though $\chi_f$ might be somewhat confusable with $\chi_b$.}

\begin{definition}[Hypergraph, \cite{scheinerman2011fractional}]  \label{def:hypergraph}
\Rarxiv{A hypergraph $\Hc$ consists of a vertex set $V(\Hc)$ and a hyperedge set $\Ec(\Hc)$, which is a family of subsets of $V(\Hc)$}. 
%A hypergraph $\Hc$ is a pair $(V,\Ec)$, where $V$ is a finite set, called the vertex set of $\Hc$, and $\Ec$ is a family of subsets of $V$, called the hyperedge set of $\Hc$. 
%The sets $V$ and $\Ec$ may also be denoted by $V(\Hc)$ and $\Ec(\Hc)$, respectively. 
\end{definition}

\Rarxiv{It can be seen that every graph is a special hypergraph whose hyperedges are all of cardinality $2$}.

%We recall the following basic definitions regarding an hypergraph \cite{scheinerman2011fractional}. 
%Similar notions can be defined analogously for the hypergraph (cf. Definition \ref{def:hypergraph}) as follows. 
For the counterparts of Definitions \ref{def:coloring}-\ref{def:fractional:chromatic} for hypergraphs, see the following. %see \cite[Sections 1.1-1.2]{scheinerman2011fractional}. 
%For definitions regarding the covering and packing of hypergraphs, see \cite[Sections 1.1-1.2]{scheinerman2011fractional}. 

%\begin{definition}[Hypergraph, \cite{scheinerman2011fractional}]
%A hypergraph $\Hc$ is a pair $(V,\Ec)$, where $V$ is a finite set, and $\Ec$ is a family of subsets of $V$. We call $V$ the vertex set of $\Hc$ and $\Ec$ the hyperedge set of $\Hc$. %We sometimes write $V$ and $\Ec$ as $V(\Hc)$ and $\Ec(\Hc)$, respectively. 
%\end{definition}

%\iffalse
\begin{definition}[Covering and covering number, \cite{scheinerman2011fractional}]  \label{def:covering}
Given a hypergraph $\Hc=(V,\Ec)$, a covering of $\Hc$ is a set of its hyperedges, $\Pcal=\{ E_1,E_2,\dots,E_m \}$ where $E_p\in \Ec,\forall p\in [m]$, such that $V=\cup_{p\in [m]}E_p$. 
The covering number of $\Hc$, denoted as $k(\Hc)$, is the smallest integer $m$ such that a covering $\Pcal=\{ E_1,E_2,\dots,E_m \}$ exists for $\Hc$. 
\end{definition}

\begin{definition}[$b$-fold covering and $b$-fold covering number, \cite{scheinerman2011fractional}]  \label{def:b:fold:covering}
Given a hypergraph $\Hc=(V,\Ec)$, a $b$-fold covering of $\Hc$ is a multiset $\Pcal=\{ E_1,E_2,\dots,E_m \}$ where $E_p\in \Ec,\forall p\in [m]$, such that every $v\in V$ is in at least $b$ sets in $\Pcal$. 
The $b$-fold covering number of $\Hc$, denoted as $k_b(\Hc)$, is the smallest integer $m$ such that a $b$-fold covering $\Pcal=\{ E_1,E_2,\dots,E_m \}$ exists for $\Hc$.
\end{definition}

\begin{definition}[Fractional Covering number, \cite{scheinerman2011fractional}]  \label{def:fractional:covering:number}
Given a hypergraph $\Hc=(V,\Ec)$, the fractional covering number $k_{f}(\Hc)$ is defined as 
\begin{align*}
k_{f}(\Hc)=\inf_{b}\frac{k_b(\Hc)}{b}=\lim_{b\to \infty}\frac{k_b(\Hc)}{b},
\end{align*}
where the second equality follows from the subadditivity of $k_b(\Hc)$ in $b$ and Fekete's Lemma \cite{fekete1923lemma}. 
\end{definition}

%\Rarxiv{packing; packing number; fractional packing number}

%\fi

%\Roy{vertex transitive}

%For a more detailed introduction on basic graph-theoretic notions, we refer the reader to \cite{scheinerman2011fractional}. 

%\iffalse
%%%%%%%%%%%%%%%%%%%%%%%%%%%%%%%%%%%%%%%%%%%%%
\section{Proof of Lemma \ref{lem:maxl:merging:consequence}}  \label{app:proof:lem:maxl:merging:consequence}

\begin{IEEEproof}
%Given any $P_{Y|X^t}$, merging to generate $P'_{Y|X^t}$. $\forall y\in \Yc'$, there exists some $T\in \Ic_{\rm max}(\Gamma_t)$ such that $\Xc^t(y)\subseteq T$ and also that $\Xc^t(\yt)\not \subseteq T$ for any $\yt\neq y\in \Yc'$. 
Consider an arbitrary $P_{Y|X^t}$. 
We keep merging any two mergeable codewords till we reach some mapping scheme $P_{Y'|X^t}$ with alphabet $\Yc'$ such that any two codewords $y_1,y_2\in \Yc'$ are not mergeable. 
According to Lemma \ref{lem:maxl:merging}, the leakage induced by $P_{Y'|X^t}$ is always no larger than that by $P_{Y|X^t}$. 
Hence, to prove the lemma, it suffices to show that there exists some mapping scheme $Q_{\Yt|X^t}$ with code alphabet ${\tilde \Yc}=\Ic_{\rm max}(\Gamma_t)$ such that the leakage induced by $Q_{\Yt|X^t}$ is no larger than that by $P_{Y'|X^t}$, i.e., 
\begin{align}
L^t(Q_{\Yt|X^t})\le L^t(P_{Y'|X^t}).  \label{eq:maxl:merging:consequence:goal}
\end{align}
%different codewords $y^1,y^2\in \Yc$ satisfying $\Xc^t(y^1)\cup \Xc^t(y^2)\subseteq T$ for some 

%\Roy{revise}
To show \eqref{eq:maxl:merging:consequence:goal}, we construct $Q_{\Yt|X^t}$ as follows. 
For every codeword $y\in \Yc'$ of the mapping $P_{Y'|X^t}$, there exists some $T\in \Ic_{\rm max}(\Gamma_t)$ such that 
\begin{align}\label{eq:maxl:merging:consequence:split}
\begin{split}
\Xc^t_{P_{Y'|X^t}}(y)&\subseteq T, \\
\Xc^t_{P_{Y'|X^t}}(y')&\not \subseteq T, \quad  \forall y'\in \Yc'\setminus \{y\},
\end{split}
\end{align}
since otherwise $y$ and $y'$ are mergeable. 
%$\Xc^t_{P_{Y'|X^t}}(y)\subseteq T$, and at the same time that $\Xc^t_{P_{Y'|X^t}}(y')\not \subseteq T$ for any other realization $y'\in \Yc'\setminus \{y\}$ (otherwise $y$ and $y'$ are mergeable). 
%Note that for every $y\in \Yc'$, there may be more than one $T\in \Ic_{\rm max}(\Gamma_t)$ that satisfied \eqref{eq:maxl:merging:consequence:split}. 
%Set ${\tilde \Yc}$ to be a subset of $\Ic_{\rm max}(\Gamma_t)$ such that for every $y\in \Yc'$ there exists one and only one $T\in {\tilde \Yc}$ satisfying \eqref{eq:maxl:merging:consequence:split}. 
Hence, it can be verified that there exists some ${\tilde \Yc}\subseteq \Ic_{\rm max}(\Gamma_t)$ such that for every $y\in \Yc'$ there exists one and only one $T\in {\tilde \Yc}$ satisfying \eqref{eq:maxl:merging:consequence:split}. 
%Then we can see that there is a one-to-one mapping between the two codebooks $\Yc'$ and ${\tilde \Yc}$. 
For every $T\in {\tilde \Yc}$, let $y(T)$ to be the unique codeword in $\Yc'$ such that $\Xc^t_{P_{Y'|X^t}}(y(T))\subseteq T$. 
Then, for any $x^t\in \Xc^t$, $T\in {\tilde \Yc}$, set 
\begin{align*}
Q_{\Yt|X^t}(T|x^t)=
%\begin{cases}
P_{Y'|X^t}(y(T)|x^t).
%\end{cases}
\end{align*}
It can be easily verified that $Q_{\Yt|X^t}$ is a valid mapping scheme. 
%As for any $x^t\in \Xc^t$, 
%\begin{align*}
%\sum_{T\in {\tilde \Yc}} Q_{\Yt|X^t}(T|x^t)
%&=\sum_{T\in {\tilde \Yc}} P_{Y'|X^t}(y(T)|x^t)  \\
%&=\sum_{y\in \Yc'} P_{Y'|X^t}(y|x^t)=1,
%\end{align*}
%we can see that $Q_{\Yt|X^t}$ is a 
%\Roy{revise}

Then, we have
\begin{align*}
L^t(Q_{\Yt|X^t})&=\log \sum_{T\in {\tilde \Yc}} \max_{x^t\in \Xc^t} Q_{\Yt|X^t}(T|x^t)  \\
&=\log \sum_{y(T):T\in {\tilde \Yc}} \max_{x^t\in \Xc^t} P_{Y'|X^t}(y(T)|x^t)  \\
&=L^t(P_{Y'|X^t}),
\end{align*}
which indicates \eqref{eq:maxl:merging:consequence:goal}, thus completing the proof. 
\end{IEEEproof}

%%%%%%%%%%%%%%%%%%%%%%%%%%%%%%%%%%%%%%%%%%%%%
\section{Proof of \eqref{eq:proof:thm:maxl:key}}  \label{app:eq:proof:thm:maxl:key}

We first prove $\eta \le 1/\chi_f(\Gamma_t)$. 
As $\eta$ is the solution to \eqref{eq:maxl:hard:distortion:optimization}, there exists some $0\le \kappa_T\le 1,T\in \Ic_{\rm max}(\Gamma_t)$ so that 
\begin{align}
&\min_{x^t\in \Xc^t} \sum_{T\in \Ic_{\rm max}(\Gamma_t):x^t\in T} \kappa_T = \eta,  \label{eq:proof:thm:maxl:eta:less:1}  \\
&\sum_{T\in \Ic_{\rm max}(\Gamma_t)} \kappa_T=1.  \label{eq:proof:thm:maxl:eta:less:2}
\end{align}
Construct $\lambda_T=\kappa_T/\eta$ for every $T\in \Ic_{\rm max}(\Gamma_t)$. 
%
%It can be verified by contradiction that $(\lambda_T: T\in \Ic_{\rm max}(\Gamma_t))$ satisfies the constraint in \eqref{eq:maxl:chif:optimization:constraint:nonnegative}. For more details, see the full version of this paper \cite{isit:2021:arxiv}. 
%
%\iffalse
We show that $\lambda_T\in [0,1]$ for any $T\in \Ic_{\rm max}(\Gamma_t)$. As $\kappa_T\ge 0$ and it can be easily verified that $\eta>0$, it is obvious that $\lambda_T\ge 0$. 
In the following we show that $\lambda_T\le 1$, which is equivalent to showing that $\kappa_T\le \eta$, by contradiction as follows. 
Let $x^t_{\Delta}$ be the vertex in $\Xc^t$ that achieves the minimum in $\min_{x^t\in \Xc^t} \sum_{T\in \Ic_{\rm max}(\Gamma_t):x^t\in T} \kappa_T$. Thus, we have $\eta=\sum_{T\in \Tc_{\Delta}}\kappa_T$, where $\Tc_{\Delta}=\{ T\in \Ic_{\rm max}(\Gamma):x^t_{\Delta}\in T \}$ denotes the set of maximal independent sets containing $x^t_{\Delta}$. 
Assume there exists some $T_0\in \Ic_{\rm max}(\Gamma_t)$ such that $\kappa_{T_0}> \eta$. Clearly, $T_0\notin \Tc_{\Delta}$, or equivalently, $x^t_{\Delta}\notin T_0$. 
We construct $\kappa'_T,T\in \Ic_{\rm max}(\Gamma_t)$ as: 
\begin{align}  \label{eq:proof:thm:maxl:contradiction:kappa:dash}
\kappa'_T=
\begin{cases}
\eta+\frac{\kappa_{T_0}-\eta}{|\Ic_{\rm max}(\Gamma_t)|},\qquad \enskip \text{if $T=T_0$},  \\
%\kappa_T+\frac{\kappa_{T_0}-\eta}{2|\Tc_{\Delta}|},\quad \text{if $T\in \Tc_{\Delta}$},  \\
\kappa_T+\frac{\kappa_{T_0}-\eta}{|\Ic_{\rm max}(\Gamma_t)|}, \qquad \text{otherwise}.
\end{cases}
\end{align}
%\begin{align}  \label{eq:proof:thm:maxl:contradiction:kappa:dash}
%\kappa'_T=
%\begin{cases}
%\frac{\eta+\kappa_{T_0}}{2},\quad \text{if $T=T_0$},  \\
%\kappa_T+\frac{\kappa_{T_0}-\eta}{2|\Tc_{\Delta}|},\quad \text{if $T\in \Tc_{\Delta}$},  \\
%\kappa_T, \quad \text{otherwise}.
%\end{cases}
%\end{align}
To verify that $\kappa'_T,T\in \Ic_{\rm max}(\Gamma)$ satisfies the constraint in \eqref{eq:maxl:hard:distortion:optimization:constraint}, we have 
\begin{align*}
&\sum_{T\in \Ic_{\rm max}(\Gamma_t)}\kappa'_T  \\
&=\kappa'_{T_0}+\sum_{T\in \Ic_{\rm max}(\Gamma_t):T\neq T_0}\kappa'_T  \\
&=(\eta+\frac{\kappa_{T_0}-\eta}{|\Ic_{\rm max}(\Gamma_t)|}) + \sum_{T\in \Ic_{\rm max}(\Gamma_t):T\neq T_0}(\kappa_T+\frac{\kappa_{T_0}-\eta}{|\Ic_{\rm max}(\Gamma_t)|})  \\
&=\eta+|\Ic_{\rm max}(\Gamma_t)|\cdot \frac{\kappa_{T_0}-\eta}{|\Ic_{\rm max}(\Gamma_t)|}+\sum_{T\in \Ic_{\rm max}(\Gamma_t):T\neq T_0}\kappa_T  \\
&=\kappa_{T_0}+\sum_{T\in \Ic_{\rm max}(\Gamma_t):T\neq T_0}\kappa_T=1,
\end{align*}
where the second equality follows from \eqref{eq:proof:thm:maxl:contradiction:kappa:dash}. 
%\begin{align*}
%&\sum_{T\in \Ic_{\rm max}(\Gamma_t)}\kappa'_T  \\
%%%
%&=\sum_{T\in \Ic_{\rm max}(\Gamma_t):T\notin \Tc_{\Delta},T\neq T_0}\kappa_T+\kappa'_{T_0}+\sum_{T\in \Tc_{\Delta}}\kappa'_T  \\
%%%
%&=1-\eta-\kappa_{T_0}+\frac{\eta+\kappa_{T_0}}{2}+\sum_{T\in \Tc_{\Delta}}\kappa_T+|\Tc_{\Delta}|\cdot \frac{\kappa_{T_0}-\eta}{2|\Tc_{\Delta}|}  \\
%%%
%&=1-\eta-\kappa_{T_0}+\frac{\eta+\kappa_{T_0}}{2}+\eta+\frac{\kappa_{T_0}-\eta}{2}  \\
%%%
%&=1,
%\end{align*}
%where the second equality follows from \eqref{eq:proof:thm:maxl:contradiction:kappa:dash} and the third equality follows from the fact that $\eta=\sum_{T\in \Tc_{\Delta}}\kappa_T$. 
It can also be verified that for any $T\in \Ic_{\rm max}(\Gamma_t)$, $\kappa'_T\in [0,1]$, thus satisfying \eqref{eq:maxl:hard:distortion:optimization:constraint:nonnegative}. 
So $(\kappa'_T: T\in \Ic_{\rm max}(\Gamma_t))$ is a valid assignment satisfying the constraints in the optimization problem \eqref{eq:maxl:hard:distortion:optimization}. 
%Therefore, $(\kappa'_T,T\in \Ic_{\rm max}(\Gamma_t))$ is a valid assignment of BLA. 
%
%
Note that we have $\kappa'_T> \kappa_T$ for any $T\in \Ic_{\rm max}(\Gamma_t)\setminus T_0$, and $\kappa'_{T_0}< \kappa_{T_0}$. 
Consider any $x^t\in \Xc^t$. 
If $x^t\notin T_0$, then we have 
\begin{align*}
&\sum_{T\in \Ic_{\rm max}(\Gamma_t):x^t\in T}\kappa'_T  \\
&> \sum_{T\in \Ic_{\rm max}(\Gamma_t):x^t\in T}\kappa_T\ge \min_{\xt^t\in \Xc^t} \sum_{T\in \Ic_{\rm max}(\Gamma_t):\xt^t\in T}\kappa_T=\eta
\end{align*}
%then $\sum_{T\in \Ic_{\rm max}(\Gamma_t):x^t\in T}\kappa'_T$ is strictly larger than $\sum_{T\in \Ic_{\rm max}(\Gamma_t):x^t\in T}\kappa_T$, which is in turn no smaller than $\eta$. 
If $x^t\in T_0$, then we have 
\begin{align*}
\sum_{T\in \Ic_{\rm max}(\Gamma_t):x^t\in T}\kappa'_T\ge \kappa'_{T_0}=\eta+\frac{\kappa_{T_0}-\eta}{|\Ic_{\rm max}(\Gamma_t)|}>\eta. 
\end{align*}
%where the last inequality follows from the assumption that $\eta<\kappa_{T_0}$. 
%Therefore, we can conclude that for any $x^t\in \Xc^t$, $\sum_{T\in \Ic_{\rm max}(\Gamma_t):x^t\in T}\kappa'_T\ge \eta$
Hence, we can conclude that with $(\kappa'_T: T\in \Ic_{\rm max}(\Gamma_t))$, the objective value in \eqref{eq:maxl:hard:distortion:optimization:objective} is strictly larger than $\eta$, 
%as
%\begin{align*}
%
%\end{align*}
which contradicts the fact that $\eta$ is the solution to optimization problem \eqref{eq:maxl:hard:distortion:optimization}. 
%\Roy{not correct yet, as $\ge \eta$ won't lead to contradiction;to correct, we need to assign a fraction of $\kappa'_{T_0}$ to $\kappa_T, T\in \Ic_{\rm max}(\Gamma_t)\setminus \Tc_{\Delta}\setminus \{T_0\}$, while keeping $\kappa'_{T_0}$ strictly larger than $\eta$.}
Therefore, the assumption that there exists some $T_0\in \Ic_{\rm max}(\Gamma_t)$ such that $\kappa_{T_0}> \eta$ must not be true, and subsequently, for every $T\in \Ic_{\rm max}(\Gamma_t)$, $\lambda_T\le 1$. In conclusion, we know that $(\lambda_T: T\in \Ic_{\rm max}(\Gamma_t))$ satisfies the constraint in \eqref{eq:maxl:chif:optimization:constraint:nonnegative}. 
%\fi
%
%
%\Roy{the above blue parts goes to ArXiv.}
%
%By \eqref{eq:proof:thm:maxl:eta:less:1}, we know that $(\lambda_T: T\in \Ic_{\rm max}(\Gamma_t))$ satisfies the constraints in \eqref{eq:maxl:chif:optimization:constraint} as for any $x^t\in \Xc^t$, 
For any $x^t\in \Xc^t$, by \eqref{eq:proof:thm:maxl:eta:less:1}, we have 
\begin{align*}
\sum_{\substack{T\in \Ic_{\rm max}(\Gamma_t):\\x^t\in T}} \lambda_T
\ge \frac{1}{\eta} \min_{v^t\in \Xc^t} \sum_{\substack{T\in \Ic_{\rm max}(\Gamma_t):\\v^t\in T}} \kappa_T=1,
\end{align*}
and thus we know that $(\lambda_T: T\in \Ic_{\rm max}(\Gamma_t))$ satisfies the constraints in \eqref{eq:maxl:chif:optimization:constraint}. 
Therefore, $(\lambda_T: T\in \Ic_{\rm max}(\Gamma_t))$ is a valid assignment satisfying the constraints in the optimization problem \eqref{eq:maxl:chif:optimization}, with which the objective in \eqref{eq:maxl:chif:optimization:objective} becomes 
\begin{align*}
\sum_{T\in \Ic_{\rm max}(\Gamma_t)} \lambda_T=\frac{1}{\eta} \sum_{T\in \Ic_{\rm max}(\Gamma_t)} \kappa_T=\frac{1}{\eta},
\end{align*}
where the second equality follows from \eqref{eq:proof:thm:maxl:eta:less:2}. 
Since $\chi_f(\Gamma_t)$ is the solution to the optimization problem \eqref{eq:maxl:chif:optimization}, 
we can conclude that $\chi_f(\Gamma_t)\le 1/\eta,$ which is equivalent to %$$\eta\le 1/\chi_f(\Gamma_t). $$
\begin{align}
\eta\le 1/\chi_f(\Gamma_t).  \label{eq:proof:thm:maxl:eta:less}
\end{align}

The opposite direction $\eta\ge 1/\chi_f(\Gamma_t)$ 
%\begin{align}
%\eta\ge 1/\chi_f(\Gamma_t)  \label{eq:proof:thm:maxl:eta:greater} %(i.e., $\eta\ge 1/\chi_f(\Gamma_t)$) 
%\end{align}
can be proved in a similar manner as follows. 
%\iffalse
As $\chi_f(\Gamma_t)$ is the solution to the optimization problem \eqref{eq:maxl:chif:optimization}, there exists some $0\le \lambda_T\le 1, T\in \Ic_{\rm max}(\Gamma_t))$ such that 
\begin{align}
&\sum_{T\in \Ic_{\rm max}(\Gamma_t)} \lambda_T=\chi_f(\Gamma_t),  \label{eq:proof:thm:maxl:eta:greater:1}  \\
&\sum_{T\in \Ic_{\rm max}(\Gamma_t):x^t\in T} \lambda_T\ge 1, \quad \forall x^t\in \Xc^t.  \label{eq:proof:thm:maxl:eta:greater:2}
\end{align}
Construct $\kappa_T=\lambda_T/\chi_f(\Gamma_t)$ for any $T\in \Ic_{\rm max}(\Gamma_t)$. 
We know that $(\kappa_T:T\in \Ic_{\rm max(\Gamma_t)})$ satisfies the constraints in \eqref{eq:maxl:hard:distortion:optimization:constraint:nonnegative} due to the simple fact that the factional chromatic number of any graph is no less than $1$. 
By \eqref{eq:proof:thm:maxl:eta:greater:1}, we know that $(\kappa_T,T\in \Ic_{\rm max}(\Gamma_t))$ satisfy the constraint in \eqref{eq:maxl:hard:distortion:optimization:constraint} as 
\begin{align*}
\sum_{T\in \Ic_{\rm max}(\Gamma_t)} \kappa_T=\frac{1}{\chi_f(\Gamma_t)} \sum_{T\in \Ic_{\rm max}(\Gamma_t)} \lambda_T=1.
\end{align*}
Therefore, $(\kappa_T: T\in \Ic_{\rm max}(\Gamma_t))$ is a valid assignment satisfying the constraints in the optimization problem \eqref{eq:maxl:hard:distortion:optimization}, with which we have 
\begin{align*}
&\min_{x^t\in \Xc^t} \sum_{T\in \Ic_{\rm max}(\Gamma_t):x^t\in T} \kappa_T  \\
&=\frac{1}{\chi_f(\Gamma_t)}\min_{x^t\in \Xc^t} \sum_{T\in \Ic_{\rm max}(\Gamma_t):x^t\in T} \lambda_T\ge \frac{1}{\chi_f(\Gamma_t)}, 
\end{align*}
where the inequality follows from \eqref{eq:proof:thm:maxl:eta:greater:2}. 
That is, the objective in \eqref{eq:maxl:hard:distortion:optimization:objective} is no smaller than $1/\chi_f(\Gamma_t)$. Since $\eta$ is the solution to the optimization problem \eqref{eq:maxl:hard:distortion:optimization}, we can conclude that 
\begin{align}
\eta\ge \frac{1}{\chi_f(\Gamma_t)}.  \label{eq:proof:thm:maxl:eta:greater}
\end{align}
%\fi
Combining \eqref{eq:proof:thm:maxl:eta:less} and \eqref{eq:proof:thm:maxl:eta:greater} yields \eqref{eq:proof:thm:maxl:key}.

%%%%%%%%%%%%%%%%%%%%%%%%%%%%%%%%%%%%%%%%%%%%%
\section{Proof of Theorem \ref{thm:multiple:guesses}}  \label{app:proof:thm:multiple:guesses}

\begin{IEEEproof}
The upper bound comes immediately from Theorem \ref{thm:maxl} and Lemma \ref{lem:multiple:guesses:less:than:maxl}. It remains to show the lower bound. 

Consider any $t$, any $P_X$, and any valid $P_{Y|X^t}$. 
We have 
\begin{align}
&\sum_{y\in \Yc} \max_{K\subseteq \Xc^t:|K|\le g(t)} \sum_{x^t\in K}P_{X^t,Y}(x^t,y)  \nonumber  \\
&\stackrel{(a)}{=} \sum_{y\in \Yc} \max_{K\subseteq \Xc^t(y):|K|\le g(t)} \sum_{x^t\in K}P_{X^t,Y}(x^t,y)  \nonumber  \\
&\ge \sum_{y\in \Yc} \frac{ \sum_{K\subseteq \Xc^t(y):|K|=g(t)^-} \sum_{x^t\in K}P_{X^t,Y}(x^t,y) }{|\{ K\subseteq \Xc^t(y):|K|=g(t)^- \}|}   \nonumber  \\
&\stackrel{(b)}{=} \sum_{y\in \Yc} \frac{ \binom{|\Xc^t(y)|-1}{g(t)^--1}\sum_{x^t\in \Xc^t(y)}P_{X^t,Y}(x^t,y) }{\binom{|\Xc^t(y)|}{g(t)^-}}  \nonumber  \\
&=\sum_{y\in \Yc} \frac{g(t)^-}{|\Xc^t(y)|} P_Y(y)  \nonumber  \\
%%
%&\stackrel{(b)}{=} \sum_{y\in \Yc:g(t)\le |\Xc^t(y)|} \frac{g(t)}{|\Xc^t(y)|} P_Y(y) + \sum_{y\in \Yc:g(t)>|\Xc^t(y)|} P_Y(y)  \nonumber  \\
%%
&\stackrel{(c)}{\ge} \frac{g(t)}{\alpha(\Gamma_t)} \sum_{y\in \Yc} P_Y(y)=\frac{g(t)}{\alpha(\Gamma_t)}, \nonumber% \sum_{y\in \Yc} P_Y(y)=\frac{g(t)}{\alpha(\Gamma_t)},
\end{align}
%where $g(t)^-=\min\{g(t),|\Xc^t(y)|\}$, and the last inequality follows from the fact that for any $y\in \Yc$, we always have $|\Xc^t(y)|\le \alpha(\Gamma_t)$. 
where $g(t)^-=\min\{g(t),|\Xc^t(y)|\}$, and 
\begin{itemize}
\item (a) follows from the fact that $P_{X^t,Y}(x^t,y)=0$ for any $x^t\notin \Xc^t(y)$ according to \eqref{eq:notation:conditional:support}; %Definition \ref{def:notation:conditional:support}; 
\item (b) follows from that each $x^t\in \Xc^t(y)$ appears in exactly $\binom{|\Xc^t(y)|-1}{g(t)^--1}$ subsets of $\Xc^t(y)$ of size $g(t)^-$;
\vspace{0.5mm}
\item (c) follows from the fact for any $y\in \Yc$, we always have $|\Xc^t(y)|\le \alpha(\Gamma_t)$ as a direct consequence of \eqref{eq:confusion:requirement} %specified in \eqref{eq:confusion:requirement:alpha} 
and thus 
%follows from that we always have $\frac{g(t)^-}{|\Xc^t(y)|}\ge \frac{g(t)}{\alpha(\Gamma_t)}$ for every $y\in \Yc$, which can be proved based on the fact that $|\Xc^t(y)|\le \alpha(\Gamma_t)$ for every $y\in \Yc$ as: 
\begin{enumerate}
\item if $g(t)\le |\Xc^t(y)|$, then $\frac{g(t)^-}{|\Xc^t(y)|}=\frac{g(t)}{|\Xc^t(y)|}\ge \frac{g(t)}{\alpha(\Gamma_t)}$, 
\item otherwise we have $g(t)>|\Xc^t(y)|$ and $\frac{g(t)^-}{|\Xc^t(y)|}=1\ge \frac{g(t)}{\alpha(\Gamma_t)}$, where the last inequality is due to the assumption that $g(t)\le \alpha(\Gamma_t)$. 
\end{enumerate}
\end{itemize}

Therefore, we have
\begin{align}
\Lc_g
&=\lim_{t\to \infty} \frac{1}{t} \inf_{\substack{P_{Y|X^t}: \\ \Xc^t(y)\in \Ic(\Gamma_t),\forall y\in \Yc}}  \nonumber  \\
&\qquad \qquad \log \max_{P_X} \frac{\sum\limits_{y\in \Yc}\max\limits_{K\subseteq \Xc^t:|K|\le g(t)}\sum\limits_{x^t\in K}P_{X^t,Y}(x^t,y)}{\max\limits_{\substack{K\subseteq \Xc^t:|K|\le g(t)}}\sum_{x^t\in K} P_{X^t}(x^t)}  \nonumber  \\
&\ge \lim_{t\to \infty} \frac{1}{t}\log \max_{P_X} \frac{\frac{g(t)}{\alpha(\Gamma_t)}}{\max\limits_{K\subseteq \Xc^t:|K|\le g(t)} \sum_{x^t\in K} P_{X^t}(x^t)}  \nonumber  \\
&\ge \lim_{t\to \infty} \frac{1}{t}\log \max_{P_X} \frac{\frac{g(t)}{\alpha(\Gamma_t)}}{g(t)\cdot \max_{x^t\in \Xc^t}P_{X^t}(x^t)}  \nonumber  \\
&\stackrel{(c)}{=} \lim_{t\to \infty} \frac{1}{t}\log \frac{|\Xc^t|}{\alpha(\Gamma_t)}  \nonumber  \\
&\stackrel{(d)}{=} \log \frac{|\Xc|}{\alpha(\Gamma)}, \nonumber
\end{align}
where (c) follows from the fact that 
\begin{align*}
\max_{P_X} \frac{1}{\max_{x^t\in \Xc^t} P_{X^t}(x^t)}\le \frac{1}{\frac{\sum_{x^t\in \Xc^t} P_{X^t}(x^t)}{|\Xc^t|}}=|\Xc^t|,
\end{align*}
which holds with equality if and only if $X$ is uniformly distributed over $\Xc$, 
and (d) follows from the fact that $|\Xc^t|=|\Xc|^t$ and $\alpha(\Gamma_t)=\alpha(\Gamma^{\lor t})=\alpha(\Gamma)^t$. 
%the latter of which is a direct consequence of Lemma \ref{}. \Roy{reference}
\end{IEEEproof}

%%%%%%%%%%%%%%%%%%%%%%%%%%%%%%%%%%%%%%%%%%%%%
\section{Proof of  Proposition \ref{prop:multiple:guesses:equal:to:maxl:small:g}}  \label{app:proof:prop:multiple:guesses:equal:to:maxl:small:g}

\begin{IEEEproof}
We write out $L_g^t(P_{Y|X^t})$ defined in \eqref{eq:def:multiple:guesses:any:t:mapping} as 
\begin{align*}
L_g^t(P_{Y|X^t})=\log \max_{P_X} \frac{\sum\limits_{y\in \Yc}\max\limits_{K\subseteq \Xc^t:|K|\le g(t)}\sum\limits_{x^t\in K}P_{X^t,Y}(x^t,y)}{\max\limits_{\substack{K\subseteq \Xc^t:|K|\le g(t)}}\sum_{x^t\in K} P_{X^t}(x^t)}.
\end{align*}

We consider the fraction in the above equality. The numerator can be bounded as
%We consider the fraction in \eqref{eq:def:multiple:guesses:any:t:mapping}. The numerator can be bounded as 
\begin{align*}
\sum_{y\in \Yc} \max_{x^t\in \Xc^t} P_{X^t,Y}(x^t,y)&\le \sum\limits_{y\in \Yc}\max\limits_{\substack{K\subseteq \Xc^t:\\|K|\le g(t)}}\sum\limits_{x^t\in K}P_{X^t,Y}(x^t,y)  \\
&\le g(t)\cdot \sum_{y\in \Yc} \max_{x^t\in \Xc^t} P_{X^t,Y}(x^t,y),
\end{align*}
and the denominator can be bounded as 
\begin{align*}
\max\limits_{x^t\in \Xc^t} P_{X^t}(x^t)
&\le \max\limits_{\substack{K\subseteq \Xc^t:|K|\le g(t)}}\sum_{x^t\in K} P_{X^t}(x^t)  \\
&\le g(t)\cdot \max\limits_{x^t\in \Xc^t} P_{X^t}(x^t)
\end{align*}

Therefore, we have 
\begin{align*}
\Lc_g
&=\lim_{t\to \infty} \frac{1}{t} \inf_{\substack{P_{Y|X^t}: \\ \Xc^t(y)\in \Ic_{\rm max}(\Gamma_t),\forall y\in \Yc}} L_g^t(P_{Y|X^t})  \\
&\ge \lim_{t\to \infty} \frac{1}{t} \inf_{\substack{P_{Y|X^t}: \\ \Xc^t(y)\in \Ic_{\rm max}(\Gamma_t),\forall y\in \Yc}}  \\
&\qquad \qquad \log \max_{P_X} \frac{\sum\limits_{y\in \Yc} \max\limits_{x^t\in \Xc^t} P_{X^t,Y}(x^t,y)}{g(t)\cdot \max\limits_{x^t\in \Xc^t} P_{X^t}(x^t)}  \\
&=\Lc-\lim_{t\to \infty}\frac{1}{t}\log g(t)=\Lc,
\end{align*}
and
\begin{align*}
\Lc_g
&=\lim_{t\to \infty} \frac{1}{t} \inf_{\substack{P_{Y|X^t}: \\ \Xc^t(y)\in \Ic_{\rm max}(\Gamma_t),\forall y\in \Yc}} L_g^t(P_{Y|X^t})  \\
&\le \lim_{t\to \infty} \frac{1}{t} \inf_{\substack{P_{Y|X^t}: \\ \Xc^t(y)\in \Ic_{\rm max}(\Gamma_t),\forall y\in \Yc}}  \\
&\qquad \qquad \log \max_{P_X} \frac{g(t)\cdot \sum\limits_{y\in \Yc} \max\limits_{x^t\in \Xc^t} P_{X^t,Y}(x^t,y)}{\max\limits_{x^t\in \Xc^t} P_{X^t}(x^t)}  \\
&=\Lc+\lim_{t\to \infty}\frac{1}{t}\log g(t)=\Lc.
\end{align*}
Combining the above results completes the proof.  
%\begin{align*}
%&\Lc^*(\Gamma,g)  \\
%%%
%&=\lim_{t\to \infty} \inf_{\substack{P_{Y|X^t}: \\ \Xc^t(y)\in \Ic_{\rm max}(\Gamma_t),\forall y\in \Yc}} \log \max_{P_X} \frac{\sum\limits_{y\in \Yc}\max\limits_{K\subseteq \Xc^t:|K|\le g(t)}\sum\limits_{x^t\in K}P_{X^t,Y}(x^t,y)}{\max\limits_{\substack{K\subseteq \Xc^t:|K|\le g(t)}}\sum_{x^t\in K} P_{X^t}(x^t)}  \\
%%%
%&\ge \lim_{t\to \infty} \inf_{\substack{P_{Y|X^t}: \\ \Xc^t(y)\in \Ic_{\rm max}(\Gamma_t),\forall y\in \Yc}} \log \max_{P_X} \frac{\sum\limits_{y\in \Yc} \max\limits_{x^t\in \Xc^t} P_{X^t,Y}(x^t,y)}{g(t)\cdot \big( \max\limits_{x^t\in \Xc^t} P_{X^t}(x^t) \big)}  \\
%%%
%&=
%\end{align*}
\end{IEEEproof}

%%%%%%%%%%%%%%%%%%%%%%%%%%%%%%%%%%%%%%%%%%%%%
\section{Proof of Theorem \ref{thm:one:approximate:guess}}  \label{app:proof:thm:one:approximate:guess}

The upper bound above immediately follows from Theorem \ref{thm:maxl} and Lemma \ref{lem:one:approximate:guess:less:than:maxl}. It remains to show the lower bound, whose proof relies on the following graph-theoretic lemmas. 

%\begin{remark}
%The upper bound comes from the fact that $\Lc(\Gamma,\Theta)\le \Lc(\Gamma)$ (\textcolor{red}{Any example $\Lc(\Gamma,\Theta)<\Lc(\Gamma)$?}). 
%\end{remark}

%{\textbf{Lemmas for proving Theorem \ref{thm:one:approximate:guess}: }}

\begin{lemma}[\cite{scheinerman2011fractional}]  \label{lem:indep}
Consider any maximal independent set $T\in \Ic_{\rm max}(\Gamma_t)$. We have $T=T_1\times T_2\times \cdots \times T_t$ for some $T_j\in \Ic_{\rm max}(\Gamma)$, $\forall j\in [t]$ (i.e., for every $j$, $T_j$ is a maximal independent set in $\Gamma$). 
\end{lemma}
%\Roy{citation to be corrected}

%%%%%%%%%%%%%%%%%
\begin{lemma}  \label{lem:neighbour}
For any $x^t=(x_1,x_2,\dots,x_t)\in V(\Theta_t)=\Xc^t$, we have $N(\Theta_t,x^t)=N(\Theta,x_1)\times N(\Theta,x_2)\times \cdots \times N(\Theta,x_t)$. Subsequently, we have $P_{X^t}(N(\Theta_t,x^t))=\prod_{j\in [t]}P_X(N(\Theta,x_j))$. 
\end{lemma}

\begin{IEEEproof}
%First, we show $N(\Theta_t,x^t)\subseteq N(\Theta,x_1)\times \cdots \times N(\Theta,x_t)$. 
Consider any $v^t=(v_1,v_2,\dots,v_t)\in N(\Theta_t,x^t)$. 
According to the definition of $\Theta_t$, %Definition \ref{def:and:product}, 
we know that for every $j\in [t]$, $v_j=x_j$ or $\{ v_j,x_j \}\in \Ec(\Theta)$. Hence, for every $j\in [t]$, $v_j\in N(\Theta,x_j)$, and thus $v^t=(v_1,\dots,v_t)\in N(\Theta,x_1)\times \cdots \times N(\Theta,x_t)$. Therefore, we know that $N(\Theta_t,x^t)\subseteq N(\Theta,x_1)\times \cdots \times N(\Theta,x_t)$. 

Now we show the opposite direction. Consider any $v^t=(v_1,v_2,\dots,v_t)\in N(\Theta,x_1)\times \cdots \times N(\Theta,x_t)$. We have $v_j\in N(\Theta,x_j)$ for every $j\in [t]$. That is, $v_j=x_j$ or $\{ v_j,x_j \}\in \Ec(\Theta)$ for every $j\in [t]$. 
Then, by the definition of $\Theta_t$, %Definition \ref{def:and:product}, 
$v^t=x^t$ or $\{ v^t,x^t \}\in \Ec(\Theta_t)$, and thus $v^t\in N(\Theta_t,x^t)$. Therefore, $N(\Theta,x_1)\times \cdots \times N(\Theta,x_t) \subseteq N(\Theta_t,x^t)$. 

In conclusion, we have $N(\Theta_t,x^t)=N(\Theta,x_1)\times \cdots \times N(\Theta,x_t)$. It remains to show that $P_{X^t}(N(\Theta_t,x^t))=\prod_{j\in [t]}P_X(N(\Theta,x_j))$. Towards that end, for every $j\in [t]$, denote $N(\Theta,x_j)$ as $\{ z_{j,1},z_{j,2},\dots,z_{j,|N(\Theta,x_j)|} \}$. 
We have 
\begin{align}
&P_{X^t}(N(\Theta_t,x^t))  \nonumber  \\
&=P_{X^t}(N(\Theta,x_1)\times \cdots \times N(\Theta,x_t))  \nonumber  \\
%&=\sum_{v^t\in N(\Theta,x_1)\times \cdots \times N(\Theta,x_t)} P_{X^t}(v^t)  \nonumber  \\
%%
&=P_{X^{t-1}}(N(\Theta,x_1)\times \cdots \times N(\Theta,x_{t-1})) \cdot P_X(z_{t,1})  \nonumber  \\
&\quad +P_{X^{t-1}}(N(\Theta,x_1)\times \cdots \times N(\Theta,x_{t-1})) \cdot P_X(z_{t,2})+\cdots  \nonumber  \\
&\quad +P_{X^{t-1}}(N(\Theta,x_1)\times \cdots \times N(\Theta,x_{t-1})) \cdot P_X(z_{t,|N(\Theta,x_t)|})  \nonumber  \\
%&=\sum_{v^{t-1}\in N(\Theta,x_1)\times \cdots \times N(\Theta,x_{t-1})} P_{X^{t-1}}(v^{t-1}) (\sum_{\ell\in [|N(\Theta,x_t)|]} P_X(z_{t,\ell}))  \nonumber  \\
%%
&=P_{X^{t-1}}(N(\Theta,x_1)\times \cdots \times N(\Theta,x_{t-1})) \cdot P_X(N(\Theta,x_t))  \nonumber  \\
&=P_{X^{t-2}}(N(\Theta,x_1)\times \cdots \times N(\Theta,x_{t-2}))  \nonumber  \\
&\quad \cdot P_X(N(\Theta,x_{t-1})) \cdot P_X(N(\Theta,x_t))  \nonumber  \\
&=\cdots  \nonumber  \\
%%
%&=\prod_{j\in [t]}(\sum_{\ell \in [|N(\Theta,x_j)|]} P_X(z_{j,\ell}))  \nonumber  \\
%%
&=\prod_{j\in [t]}P_X(N(\Theta,x_j)),  \nonumber
\end{align}
which completes the proof. 
\end{IEEEproof}

\begin{lemma}  \label{lem:intersection:indep:neighbour}
Consider any maximal independent set $T=T_1\times T_2\times \dots \times T_t\in \Ic_{\rm max}(\Gamma_t)$ and any $x^t=(x_1,x_2,\cdots,x_t)\in \Xc^t$. We have 
\begin{align*}
T\cap N(\Theta_t,x^t)=\prod_{j\in [t]} (T_j\cap N(\Theta,x_j)). 
\end{align*}
%\begin{align}
%T\cap N(\Theta_t,x^t)&=(T_1\cap N(\Theta,x_1))\times (T_2\cap N(\Theta,x_2))  \nonumber  \\
%&\qquad \times \dots \times (T_t\cap N(\Theta,x_t)).
%\end{align}
\end{lemma}

\begin{IEEEproof}
%First, we show $T\cap N(\Theta_t,x^t)\subseteq (T_1\cap N(\Theta,x_1))\times \dots \times (T_t\cap N(\Theta,x_t))$. 
Consider any $v^t=(v_1,\dots,v_t)\in T\cap N(\Theta_t,x^t)$. We know $v^t\in T$ and $v^t\in N(\Theta_t,x^t)$, which, together with Lemmas \ref{lem:indep} and \ref{lem:neighbour}, indicate that for every $j\in [t]$, $v_j\in T_j$ and $v_j\in N(\Theta,x_j)$. Hence, $v_j\in T_j\cap N(\Theta,x_j)$ for every $j\in [t]$ and thus $v^t\in (T_1\cap N(\Theta,x_1))\times \dots \times (T_t\cap N(\Theta,x_t))$. Therefore, we know that $T\cap N(\Theta_t,x^t)\subseteq \prod_{j\in [t]} (T_j\cap N(\Theta,x_j))$. 

Now we show the opposite direction. Consider any $v^t=(v_1,\dots,v_t)\in (T_1\cap N(\Theta,x_1))\times \dots \times (T_t\cap N(\Theta,x_t))$. We know that for every $j\in [t]$, $v_j\in T_j$ and $v_j\in N(\Theta,x_j)$. Hence, by Lemmas \ref{lem:indep} and \ref{lem:neighbour}, $v^t\in T$ and $v^t\in N(\Theta_t,x^t)$, and thus $v^t\in T\cap N(\Theta_t,x^t)$. Therefore, we know that $\prod_{j\in [t]} (T_j\cap N(\Theta,x_j))\subseteq T\cap N(\Theta_t,x^t)$. 

In conclusion, we have $T\cap N(\Theta_t,x^t)=\prod_{j\in [t]} (T_j\cap N(\Theta,x_j))$. 
\end{IEEEproof}

%\Roy{revise}

%%%-------------------------------- A Key Corollary
%Consider any maximal independent set $T=T_1\times T_2\times \dots \times T_t\in \Ic_{\rm max}(\Gamma_t)$. By Lemma \ref{lem:indep} we know that $T_j\in \Ic_{\rm max}(\Gamma)$, $\forall j\in [t]$. We have the following corollary. 
%
%\begin{corollary}  \label{cor:associated:hypergraph:product}
%%Consider any maximal independent set $T=T_1\times T_2\times \dots \times T_t\in \Ic_{\rm max}(\Gamma_t)$. Then, we know 
%We have $\Hc(T)=\Hc(T_1)\times \Hc(T_2) \times \dots \times \Hc(T_t)$, where for any two hypergraphs $\Hc_1$ and $\Hc_2$, the product $\Hc_1\times \Hc_2$ is defined in \cite[Section 1.6]{scheinerman2011fractional}. Moreover, we have $k_f(\Hc(T))=\prod_{j\in [t]}k_f(\Hc(T_j))$. 
%\end{corollary}
%
%The proof of the above corollary follows immediately from Lemma \ref{lem:intersection:indep:neighbour} and \cite[Theorem 1.6.1]{scheinerman2011fractional}. 

%%%%%%%%%%%%%%%%
\begin{lemma}  \label{lem:covering:indep:neighbour}
Consider any maximal independent set $T=T_1\times T_2\times \dots \times T_t\in \Ic_{\rm max}(\Gamma_t)$. 
For every $j\in [t]$, let $\Pcal_j=\{ E_{j,1},E_{j,2},\dots,E_{j,m_j} \}$ be an arbitrary $b_j$-fold covering\footnote{Recall that for any integer $b\ge 2$, any $b$-fold covering of a hypergraph is a multiset. } of the hypergraph $\Hc_1(T_j)$, where for every $i\in [m_j]$, set $E_{j,i}$ denotes the intersection of $T_j$ and the neighbors of some vertex $x_{j,i}$. That is, $E_{j,i}=T_j\cap N(\Theta,x_{j,i})$. 
Then we know that 
\begin{align*}
\Pcal&=\prod_{j\in [t]}\Pcal_j=\{ E_{1,1},\dots,E_{1,m_1} \} \times \dots \times \{ E_{t,1},\dots,E_{t,m_t} \}
\end{align*}
is a valid $\prod_{j\in [t]}b_j$-fold covering of the hypergraph $\Hc_t(T)$ with cardinality $|\Pcal|=\prod_{j\in [t]}m_j$.
\end{lemma}

\begin{IEEEproof}
The proof can be decomposed into two parts: (I) we show that every element of $\Pcal$ is a hyperedge of the hypergraph $\Hc_t(T)$; (II) we show that every vertex $x^t$ in $V(\Hc_t(T))=T$ appears in at least $\prod_{j\in [t]}b_j$ sets in $\Pcal$. 

To show part (I), without loss of generality, consider the set $E=E_{1,1}\times E_{2,1}\times \cdots \times E_{t,1}$, which is an element of $\Pcal$. 
Recall that for every $j\in [t]$, $E_{j,1}=T_j\cap N(\Theta,x_{j,1})$ for some $x_{j,1}\in \Xc$. By Lemma \ref{lem:intersection:indep:neighbour} we have 
\begin{align*}
E&=(T_1\cap N(\Theta,x_{1,1}))\times \cdots \times (T_t\cap N(\Theta,x_{t,1}))  \nonumber  \\
&=T\cap N(\Theta_t,(x_{1,1},\dots,x_{t,1})).
\end{align*}
Hence, one can see that set $E$ is indeed a hyperedge of $\Hc_t(T)$ (cf. Definition \ref{def:associated:hypergraph}). 

To show part (II), consider any $x^t=(x_1,\dots,x_t)\in T$. Since for every $j\in [t]$, $\Pcal_j$ is a $b_j$-fold covering of $\Hc_1(T_j)$, we know that vertex $x_j\in T_j$ appears in at least $b_j$ sets within $\Pcal_j$. Therefore, $x^t$ appears in at least $\prod_{j\in [t]}b_j$ sets in $\Pcal=\prod_{j\in [t]}\Pcal_j$. 
\end{IEEEproof}

\vspace{1mm}

%{\textbf{Proof of Theorem \ref{thm:one:approximate:guess}: }}
We are ready to show the lower bound in Theorem \ref{thm:one:approximate:guess}. 

\begin{IEEEproof}[Proof of Theorem \ref{thm:one:approximate:guess}]
Consider any sequence length $t$, valid mapping $P_{Y|X^t}$, and source distribution $P_X$. 

Consider any codeword $y\in \Yc$ and any maximal independent set $T=T_1\times \dots \times T_t\in \Ic_{\rm max}(\Gamma_t)$ such that $\Xc^t(y)\subseteq T$. 

For every $j\in[t]$, let $\Pcal_j=\{ E_{j,1},\dots,E_{j,m_j} \}$ be the $k_f(\Hc_1(T_j))$-achieving $b_j$-fold covering of the hypergraph $\Hc_{T_j}$. 
Note that the existence of such $\Pcal_j$ for some finite integer $b_j$ is guaranteed by the fact that $\Hc_1(T_j)$ has no exposed vertices (i.e., every vertex of $\Hc_1(T_j)$ is in at least one hyperedge of $\Hc_1(T_j)$) \cite[Corollary 1.3.2]{scheinerman2011fractional} . 

Construct $\Pcal=\prod_{j\in [t]}\Pcal_j$. By Lemma \ref{lem:covering:indep:neighbour}, we know that $\Pcal$ is a $(\prod_{j\in [t]}b_j)$-fold covering of $\Hc_t(T)$ and that $|\Pcal|=\prod_{j\in [t]}m_j$. 
Note that we have 
\begin{align}
\frac{\prod_{j\in [t]}m_j}{\prod_{j\in [t]}b_j}=\prod_{j\in [t]}k_f(\Hc_1(T_j)).  \label{eq:one:approximate:guess:covering:cardinality}
\end{align}

Recall that every element of $\Pcal$ is a hyperedge of $\Hc_t(T)$. Also recall that by Definition \ref{def:associated:hypergraph} every hyperedge $E$ of $\Hc_t(T)$ equals to $T\cap N(\Theta_t,x^t)$ for some $x^t\in \Xc^t$. 
Then we have 
\begin{align}
&\max_{x^t\in \Xc^t} \sum_{\xt^t\in N(\Theta_t,x^t)} P_{X^t,Y}(\xt^t,y)  \nonumber  \\
&\stackrel{(a)}{=} \max_{x^t\in \Xc^t} \sum_{\xt^t\in T\cap N(\Theta_t,x^t)} P_{X^t,Y}(\xt^t,y)  \nonumber  \\
&=\max_{E\in \Ec(\Hc_t(T))} \sum_{x^t\in E} P_{X^t,Y}(\xt^t,y)  \nonumber  \\
&\ge \frac{1}{|\Pcal|} \sum_{E\in \Pcal} \sum_{\xt^t\in E} P_{X^t,Y}(\xt^t,y)  \nonumber  \\
&\stackrel{(b)}{\ge} \frac{1}{\prod_{j\in [t]}m_j} (\prod_{j\in [t]}b_j) \sum_{x^t\in T}P_{X^t,Y}(x^t,y)  \nonumber  \\
&\stackrel{(c)}{=} \frac{1}{\prod_{j\in [t]}k_f(\Hc_1(T_j))} P_{Y}(y)  \nonumber  \\
%%
%&\ge \frac{P_Y(y)}{(\max_{T_0\in \Ic_{\rm max}(\Gamma)}k_f(T_0))^t}  \nonumber  \\
%%
&\ge \frac{P_Y(y)}{(\max_{T_0\in \Ic_{\rm max}(\Gamma)} k_f(\Hc_1(T_0)))^t},  \label{eq:one:approximate:guess:any:y:lower}
\end{align}
where (a) follows from the fact that $P_{X^t,Y}(x^t,y)=0$ for any $x^t\in \Xc^t\setminus T\subseteq \Xc^t\setminus \Xc^t(y)$ according to \eqref{eq:notation:conditional:support}, (b) follows from the fact that $|\Pcal|=\prod_{j\in [t]}m_j$ and that every $x^t\in V(\Hc_t(T))=T$ appears in at least $\prod_{j\in [t]}b_j$ hyperedges within $\Pcal$, and (c) follows from \eqref{eq:one:approximate:guess:covering:cardinality}. 
%and (c) follows from \eqref{eq:def:kf:gamma:theta}. 

\iffalse
For every element of $\Pcal$, simply denoted by $E$, there is a corresponding $x^t\in \Xc^t$ such that $E=T\cap N(\Theta_t,x^t)$. 
Define multiset $K(\Pcal)$ to be the collection of these $x^t$ corresponding to the elements of $\Pcal$. 
Thus $|K(\Pcal)|=|\Pcal|=\prod_{j\in [t]}m_j$. 

Then we have 
\begin{align}
&\max_{x^t\in \Xc^t} \sum_{\xt^t\in N(\Theta_t,x^t)} P_{X^t,Y}(\xt^t,y)  \nonumber  \\
%%
&=\max_{x^t\in \Xc^t} \sum_{\xt^t\in T\cap N(\Theta_t,x^t)} P_{X^t,Y}(\xt^t,y)  \nonumber  \\
%%
&\ge \frac{1}{|K(\Pcal)|} \sum_{x^t\in K(\Pcal)} \sum_{\xt^t\in T\cap N(\Theta_t,x^t)} P_{X^t,Y}(\xt^t,y)  \nonumber  \\
%%
&\stackrel{(a)}{\ge} \frac{1}{\prod_{j\in [t]}m_j} (\prod_{j\in [t]}b_j) (\sum_{x^t\in T}P_{X^t,Y}(x^t,y))  \nonumber  \\
%%
&\stackrel{(b)}{=} \frac{1}{\prod_{j\in [t]}k_f(\Hc_1(T_j))} P_{Y}(y)  \nonumber  \\
%%
%&\ge \frac{P_Y(y)}{(\max_{T_0\in \Ic_{\rm max}(\Gamma)}k_f(T_0))^t}  \nonumber  \\
%%
&\ge \frac{P_Y(y)}{(\max_{T\in \Ic_{\rm max}(\Gamma)} k_f(\Hc_1(T)))^t},  \label{eq:one:approximate:guess:any:y:lower}
\end{align}
where (a) follows from the fact that $|K(\Pcal)|=\prod_{j\in [t]}m_j$ and that every $x^t\in T$ appears in at least $\prod_{j\in [t]}b_j$ sets within $\Pcal$, 
and (b) follows from \eqref{eq:one:approximate:guess:covering:cardinality}. 
%and (c) follows from \eqref{eq:def:kf:gamma:theta}. 
\fi

Therefore, we have 
\begin{align}
&\Lc_{\Theta}  \nonumber  \\
&=\lim_{t\to \infty} \frac{1}{t} \inf_{\substack{P_{Y|X^t}: \\ \Xc^t(y)\in \Ic(\Gamma_t),\forall y\in \Yc}}  \nonumber  \\
&\qquad \qquad \log \max_{P_X} \frac{\sum\limits_{y\in \Yc} \max\limits_{x^t\in \Xc^t} \sum\limits_{\xt^t\in N(\Theta_t,x^t)} P_{X^t,Y}(\xt^t,y)}{\max\limits_{x^t\in \Xc^t} \sum\limits_{\xt^t\in N(\Theta_t,x^t)} P_{X^t}(\xt^t)}  \nonumber  \\
&\stackrel{(d)}{\ge} \lim_{t\to \infty} \frac{1}{t}\log \max_{P_X} \frac{\sum_{y\in \Yc}\frac{P_Y(y)}{(\max_{T\in \Ic_{\rm max}(\Gamma)} k_f(\Hc_1(T)))^t}}{\max_{x^t\in \Xc^t}P_{X^t}(N(\Theta_t,x^t))}  \nonumber  \\
&\stackrel{(e)}{=} \lim_{t\to \infty} \frac{1}{t}\log \max_{P_X} \frac{\sum_{y\in \Yc}\frac{P_Y(y)}{(\max_{T\in \Ic_{\rm max}(\Gamma)} k_f(\Hc_1(T)))^t}}{(\max_{x\in \Xc} P_X(N(\Theta,x)))^t}  \nonumber  \\
&=\log \frac{1}{\max\limits_{T\in \Ic_{\rm max}(\Gamma)} k_f(\Hc_1(T))} + \log \max_{P_X}\min_{x\in \Xc} \frac{1}{P_X(N(\Theta,x))}  \nonumber  \\
&=\log \frac{1}{\max\limits_{T\in \Ic_{\rm max}(\Gamma)} k_f(\Hc_1(T))} + \log \frac{1}{\min\limits_{P_X} \max\limits_{x\in \Xc} P_X(N(\Theta,x))},  \nonumber
\end{align}
where (d) follows from \eqref{eq:one:approximate:guess:any:y:lower} and (e) is due to Lemma \ref{lem:neighbour}. 

\Rarxiv{It remains to show that $$p_f(\Theta)=\frac{1}{\min_{P_X} \max_{x\in \Xc} P_X(N(\Theta,x))}, $$
or equivalently, $p_f(\Theta)=1/\tau$,  
where $\tau$ is the solution to the following optimization problem: 
\begin{subequations}  \label{eq:approximate:optimization}
\begin{align}
\text{minimize} \quad &\max_{x\in \Xc} \sum_{\xt\in N(\Theta,x)} P_X(\xt),  \label{eq:approximate:optimization:objective}  \\
\text{subject to} \quad &\sum_{x\in \Xc}P_X(x)=1,  \label{eq:approximate:optimization:constraint}  \\
&P_X(x)\in [0,1], \quad \forall x\in \Xc.  \label{eq:approximate:optimization:constraint:nonnegative}
\end{align}
\end{subequations}
Recall that $p_f(\Theta)$ denotes the fractional closed neighborhood packing number of $\Theta$, which is the solution to the following linear program \cite[Section 7.4]{scheinerman2011fractional}:  %optimization problem: 
\begin{subequations}  \label{eq:packing:optimization}
\begin{align}
\text{maximize} \quad &\sum_{x\in \Xc} \lambda(x),  \label{eq:packing:optimization:objective}  \\
\text{subject to} \quad &\sum_{\xt\in N(\Theta,x)} \lambda(\xt)\le 1, \quad \forall x\in \Xc,  \label{eq:packing:optimization:constraint}  \\
&\lambda(x)\in [0,1], \quad \forall x\in \Xc.  \label{eq:packing:optimization:constraint:nonnegative}
\end{align}
\end{subequations}
Using similar techniques to the proof of Theorem \ref{thm:maxl}, we can show that the solutions to \eqref{eq:approximate:optimization} and \eqref{eq:packing:optimization} are reciprocal to each other. That is, $p_f(\Theta)=1/\tau$, which completes the proof of Theorem \ref{thm:one:approximate:guess}. 
}
\end{IEEEproof}

%\iffalse
%%%%%%%%%%%%%%%%%%%%%%%%%%%%%%%%%%%%%%%%%%%%%
\section{Proof of Theorem \ref{thm:multiple:approximate:guesses}}  \label{app:proof:thm:multiple:approximate:guesses}

%\Rarxiv{to be revised}

%\iffalse
\begin{IEEEproof}
%When the context is clear, we may simply use $k_f$ to denote $k_f(\Gamma,\Theta)$, which is defined in \eqref{eq:def:kf:gamma:theta}. 
%$=\max_{T\in \Ic_{\rm max}(\Gamma)}k_f(\Hc(T))$. 
Throughout the proof, we use the shorthand notation 
\begin{align}
k_f=\max_{T\in \Ic_{\rm max}(\Gamma)}k_f(\Hc_1(T)).  \label{eq:multiple:approximate:guesses:kf}
\end{align} 

%\begin{lemma}  \label{lem:sigma:less:than:logkf}
%For any 
%\end{lemma}

%Now we prove Theorem \ref{thm:multiple:approximate:guesses} in the following. 

The upper bound in Theorem \ref{thm:multiple:approximate:guesses} immediately follows from Theorem \ref{thm:maxl} and Lemma \ref{lem:multiple:approximate:guesses:less:than:maxl}. 
%the fact that $\Lc^*(\Gamma,\Theta,g)\le \Lc^*(\Gamma)=\log \chi_f(\Gamma)$. 
It remains to show the lower bound. 

Consider any sequence length $t$, any source distribution $P_X$ and any valid mapping $P_{Y|X^t}$. 
%We have
%\begin{align}
%&\sum_{y\in \Yc} \max_{\substack{K\subseteq \Xc^t:\\|K|\le k(t)}} \sum_{\substack{x^t\in \cup_{\xt^t\in K}N(\Theta_t,\xt^t)}} P_{X^t,Y}(x^t,y)  \nonumber  \\
%%%
%&=\sum_{y\in \Yc} \max_{\substack{K\subseteq \Xc^t:\\|K|\le k(t)}} \sum_{\substack{x^t\in \cup_{\xt^t\in K}(N(\Theta_t,\xt^t)\cap \Xc^t(y))}} P_{X^t,Y}(x^t,y)  \nonumber,
%\end{align}
%simply due to the fact that $P_{X^t,Y}(x^t,y)=0$ for any $x^t\notin \Xc^t(y)$ according to Definition \ref{def:notation:conditional:support}. 
Consider any codeword $y\in \Yc$. 
%We have $\Xc^t(y)\in \Ic(\Gamma_t)$. 
There exists some $T\in \Ic_{\rm max}(\Gamma_t)$ such that $\Xc^t(y)\subseteq T$. 
By Lemma \ref{lem:indep}, we have $T=T_1\times T_2\times \dots \times T_t$ where $T_j\in \Ic_{\rm max}(\Gamma), \forall j\in [t]$. 
For every $j\in[t]$, let $\Pcal_j=\{ E_{j,1},\dots,E_{j,m_j} \}$ be the $k_f(\Hc_1(T_j))$-achieving $b_j$-fold covering of the hypergraph $\Hc_1(T_j)$.  
%(cf. Definition \ref{def:associated:hypergraph}). 
%Note that the existence of such $\Pcal_j$ for some finite positive integer $b_j$ is guaranteed by \cite[Corollary 1.3.2]{scheinerman2011fractional} and the fact that $\Hc(T_j)$ has no exposed vertices (i.e., every vertex of $\Hc(T_j)$ is in at least one hyperedge of $\Hc(T_j)$). 

Construct $\Pcal=\prod_{j\in [t]}\Pcal_j$. 
Then $|\Pcal|=\prod_{j\in [t]}m_j$. 
By Lemma \ref{lem:covering:indep:neighbour}, we know that $\Pcal$ is a $(\prod_{j\in [t]}b_j)$-fold covering of $\Hc_t(T)$. 
Note that $\Pcal$ is a multiset (cf. Appendix \ref{app:basic:graph:theoretic:notions}). 
Set $m=\prod_{j\in [t]}m_j$ and $b=\prod_{j\in [t]}b_j$. 
Then $\Pcal$ is a $b$-fold covering of $\Hc_t(T )$ of cardinality $m$, and we have 
\begin{align}
\frac{m}{b}=\frac{\prod_{j\in [t]}m_j}{\prod_{j\in [t]}b_j}=\prod_{j\in [t]}k_f(\Hc_1(T_j)).  \label{eq:multiple:approximate:guesses:covering:cardinality}
\end{align} 

We assume that $b\ge g(t)$ without loss of generality.\footnote{If $b<g(t)$, we can simply construct a $cb$-fold covering of $\Hc_t(T)$, $\Pcal_c$, from $\Pcal$ by repeating it $c$ times, for some sufficiently large integer $c$ such that $cb\ge g(t)$. Then the remaining proof will be based on $\Pcal_c$. } 

For every hyperedge of $\Hc_t(T)$ in the covering $\Pcal$, denoted by $E$, there is a corresponding $x^t\in \Xc^t$ such that $E=T\cap N(\Theta_t,x^t)$. 
Let $\Xc^t(\Pcal)$ denote the collection of the corresponding $x^t$ of those hyperedges in $\Pcal$. 
More precisely, define $\Xc^t(\Pcal)$ as the multiset of $x^t$ whose corresponding hyperedge $E=T\cap N(\Theta_t,x^t)$ appears in the fractional covering $\Pcal$, while setting the multiplicity of any $x^t\in \Xc^t(\Pcal)$ the same as that of its corresponding hyperedge in $\Pcal$. 
Thus $|\Xc^t(\Pcal)|=|\Pcal|=m$. 
%Note that $\Pcal$ is a multiset and so is $\Xc^t(\Pcal)$. 

Let $\Kc=\{ K\subseteq \Xc^t(\Pcal) :|K|=g(t) \}$ denote the collection of subsets of $\Xc^t(\Pcal)$ that is of cardinality $g(t)$. 
Hence $|\Kc|=\binom{m}{g(t)}$. 
Note that any $K\in \Kcal$ is also a multiset. 

For any $K\in \Kcal$, and any $v^t\in T$, let $m(K,v^t)$ denote the number of elements $x^t$ in $K$ whose neighborhood in $\Theta_t$ contains $v^t$. That is 
\begin{align*}
m(K,v^t)=|\{ x^t\in K:v^t\in N(\Theta_t,x^t) \}|. 
\end{align*} 
Then for any $K\in \Kcal$ and $v^t\in T$, we have $$0\le m(K,v^t)\le |K|=g(t). $$

% We change our notations from the sketch notes: We use here M to denote KB, B to denote B, B(v^t) to denote B_v

%\Roy{Change notations $M$ and $B(v^t)$ to $m$ and $b(v^t)$.} 

For any $v^t\in T$, it appears in at least $b$ hyperedges in $\Pcal$. Assume $v^t$ appears in $b(v^t)$ hyperedges in $\Pcal$. Thus 
\begin{align}
b(v^t)\ge b\ge g(t).  \label{eq:multiple:approximate:guesses:bvt:b:gt}
\end{align}

\Rarxiv{Also, define the shorthand notation $N(\Theta_t,K)=\cup_{x^t\in K}N(\Theta_t,x^t)$}. 

We have 
\begin{align}
&\max_{\substack{K\subseteq \Xc^t:\\|K|\le g(t)}} \sum_{\substack{x^t\in N(\Theta_t,K)}} P_{X^t,Y}(x^t,y)  \nonumber  \\
&\stackrel{(a)}{=} \max_{\substack{K\subseteq \Xc^t:\\|K|\le g(t)}} \sum_{\substack{x^t\in \cup_{\xt^t\in K}(N(\Theta_t,\xt^t)\cap T)}} P_{X^t,Y}(x^t,y)  \nonumber  \\
&\ge \frac{1}{|\Kc|} \big( \sum_{K\in \Kc} \sum_{x^t\in \cup_{\xt^t\in K}(N(\Theta_t,\xt^t)\cap T)} P_{X^t,Y}(x^t,y) \big)  \nonumber  \\
&=\frac{1}{|\Kc|} \big( \sum_{K\in \Kc} \sum_{v^t\in T:m(K,v^t)\ge 1} P_{X^t,Y}(v^t,y) \big)  \nonumber  \\
&=\frac{1}{|\Kc|} \big( \sum_{v^t\in T} P_{X^t,Y}(v^t,y) (\sum_{K\in \Kc:m(K,v^t)\ge 1}1) \big)  \nonumber  \\
&=\frac{1}{|\Kc|} \big( \sum_{v^t\in T} P_{X^t,Y}(v^t,y) (\sum_{\ell\in [g(t)]} \sum_{K\in \Kc:m(K,v^t)=\ell}1) \big)  \nonumber  \\
&=\frac{1}{|\Kc|} \big( \sum_{v^t\in T} P_{X^t,Y}(v^t,y) \sum_{\ell\in [g(t)]} \binom{b(v^t)}{\ell} \binom{m-b(v^t)}{g(t)-\ell} \big)  \nonumber  \\
&\stackrel{(b)}{=} \frac{1}{|\Kc|} \big( \sum_{v^t\in T} P_{X^t,Y}(v^t,y) (\binom{m}{g(t)}-\binom{m-b(v^t)}{g(t)}) \big)  \nonumber  \\
&\stackrel{(c)}{\ge} (1-(\frac{m-b}{m})^{g(t)}) \sum_{v^t\in T} P_{X^t,Y}(v^t,y)   \nonumber  \\
&= (1-(\frac{m-b}{m})^{g(t)}) P_Y(y)  \nonumber  \\
&\stackrel{(d)}{\ge} (1-(1-(\frac{1}{k_f})^t)^{g(t)}) P_Y(y),  \label{eq:multiple:approximate:guesses:mid}
\end{align}
where 
\begin{itemize}
\item (a) follows from the fact that $P_{X^t,Y}(x^t,y)=0$ for any $x^t\in \Xc^t\setminus T\subseteq \Xc^t\setminus \Xc^t(y)$ according to \eqref{eq:notation:conditional:support}; 
\item 
(b) can be shown 
%(a) can be shown utilizing the standard double counting technique in combinatorics. 
by considering a specific way of choosing $g(t)$ elements from a set, denoted by $M$, of cardinality $|M|=m$, which is described as follows. 
We arbitrarily pick $b(v^t)$ elements from the set $M$, the collection of which is denoted by $B$. 
Recall that $|B|=b(v^t)\ge g(t)$ as specified in \eqref{eq:multiple:approximate:guesses:bvt:b:gt}. 
We observe that to choose $g(t)$ elements from the set $M$, the number of chosen elements from the subset $B$ must be some integer $\ell$ from $0$ to $g(t)$. 
For each possible $\ell$, the number of ways in which $b(v^t)$ elements can be chosen from $M$ is $\binom{b(v^t)}{\ell} \binom{m-b(v^t)}{g(t)-\ell}$. 
Therefore, the total number of ways to choose $g(t)$ elements from $M$ is $\sum_{\ell \in [0:g(t)]} \binom{b(v^t)}{\ell} \binom{m-b(v^t)}{g(t)-\ell}$, 
which should be equal to $\binom{m}{g(t)}$. 
Therefore, we have 
$\sum_{\ell \in [g(t)]} \binom{b(v^t)}{\ell} \binom{m-b(v^t)}{g(t)-\ell}=\binom{m}{g(t)} - \binom{b(v^t)}{0} \binom{m-b(v^t)}{g(t)-0}=\binom{m}{g(t)} - \binom{m-b(v^t)}{g(t)}$; 
%As we know the total number of ways to choose $g(t)$ elements from the set $M$ should be equal to $\binom{m}{g(t)}$, we have $\binom{m}{g(t)}=\sum_{\ell \in [0:g(t)]} \binom{b(v^t)}{\ell} \binom{m-b(v^t)}{g(t)-\ell}$, which in turn indicates that 
%\begin{align*}
%&\sum_{\ell \in [g(t)]} \binom{b(v^t)}{\ell} \binom{m-b(v^t)}{g(t)-\ell}  \\
%&=\binom{m}{g(t)} - \binom{b(v^t)}{0} \binom{m-b(v^t)}{g(t)-0}  \\
%&=\binom{m}{g(t)} - \binom{m-b(v^t)}{g(t)}. 
%\end{align*}
\vspace{1mm}
\item (c) follows from the following derivation: 
\begin{align*}
&\frac{1}{|\Kcal|} (\binom{m}{g(t)} - \binom{m-b(v^t)}{g(t)})  \\
&=\frac{1}{\binom{m}{g(t)}} (\binom{m}{g(t)} - \binom{m-b(v^t)}{g(t)})  \\
&=1-\prod_{i\in [0:g(t)-1]} \frac{m-b(v^t)-i}{m-i}  \\
&\ge 1-\prod_{i\in [0:g(t)-1]} \frac{m-b(v^t)}{m}  \\
&=1-(\frac{m-b(v^t)}{m})^{g(t)}  \\
&\ge 1-(\frac{m-b}{m})^{g(t)},
\end{align*}
where the last inequality follows from \eqref{eq:multiple:approximate:guesses:bvt:b:gt};  
%where the last inequality follows from the fact that $\Pcal$ is a $b$-fold covering of $\Hc(T)$ and thus every $v^t\in T$ appears in at least $b$ sets in $\Pcal$ (i.e., $b(v^t)\ge b$); 
%%
\item (d) follows from \eqref{eq:multiple:approximate:guesses:kf} and \eqref{eq:multiple:approximate:guesses:covering:cardinality}. 
\end{itemize}

Given \eqref{eq:multiple:approximate:guesses:mid}, it remains to further lower-bound the term $$\lim_{t\to \infty} \frac{1}{t} \log (1-(1-(\frac{1}{k_f})^t)^{g(t)}). $$

Define $$\sigma=\lim_{t\to \infty}\frac{1}{t}\log g(t). $$

Due to Proposition \ref{prop:multiple:approximate:guesses:equal:to:one:approximate:guess:small:g}, it suffices to consider only the case when $\sigma>0$ and subsequently $2^{\sigma}>1$. 

Consider any positive real number $1<m< 2^{\sigma}$. 

We have $$0<\sigma-\log m=\lim_{t\to \infty}\frac{1}{t}\log \frac{g(t)}{m^t}, $$which indicates that 
\begin{align}
\lim_{t\to \infty}g(t)>\lim_{t\to \infty}m^t.  \label{eq:multiple:approximate:guesses:m:1}
\end{align}

Towards bounding $\lim_{t\to \infty} \log \frac{1}{t} (1-(1-(\frac{1}{k_f})^t)^{g(t)})$, we first show that 
\begin{align}
m< k_f,  \label{eq:multiple:approximate:guesses:m:2}
\end{align} 
which is equivalent to showing that $\sigma \le \log k_f$. Towards that end, we have 
\begin{align}
\sigma&=\lim_{t\to \infty}\frac{\log g(t)}{t}  \nonumber  \\
&\stackrel{(e)}{\le} \lim_{t\to \infty}\frac{\log \max_{T\in \Ic_{\rm max}(\Gamma_t)} k(\Hc_t(T))}{t}  \nonumber  \\
&\stackrel{(f)}{\le} \lim_{t\to \infty}\frac{\log (1+\log e_0^t) \max_{T\in \Ic_{\rm max}(\Gamma_t)}k_f(\Hc_t(T))}{t}  \nonumber  \\
&\stackrel{(g)}{=} \lim_{t\to \infty}\frac{\log (1+\log e_0^t) (k_f)^t}{t}  \nonumber  \\
&\stackrel{(h)}{=} \lim_{t\to \infty} \frac{1}{\ln 2} \frac{(1+e_0t)(k_f)^t(\ln k_f)+(k_f)^te_0}{(k_f)^t(1+e_0t)}  \nonumber  \\
&\stackrel{(i)}{=} \frac{\ln k_f}{\ln 2}=\log k_f,
\end{align}
where 
\begin{itemize}
\item (e) follows from the assumption that $g(t)\le \max_{T\in \Ic_{\rm max}(\Gamma)} k(\Hc_t(T))$; 
\item (f) follows from \cite[Lemma 1.6.4]{scheinerman2011fractional} with $e_0=\max\limits_{T_0\in \Ic_{\rm max}(\Gamma),x_0\in \Xc} |T_0\cap N(\Theta,x_0)|$; 
\item (g) follows from the fact that for any $T=T_1\times \cdots \times T_t\in \Ic_{\rm max}(\Gamma_t)$, we have $k_f(\Hc_t(T))=\prod_{j\in [t]}k_f(\Hc_1(T_j))$, which can be shown using Lemma \ref{lem:intersection:indep:neighbour} and \cite[Theorem 1.6.1]{scheinerman2011fractional}. 
%Corollary \ref{cor:associated:hypergraph:product}; 
%%
\item (h) follows from L'Hôpital's rule;
\item (i) follows from the fact that $\lim_{t\to \infty} \frac{(k_f)^te_0}{(k_f)^t(1+e_0t)}=0$. 
\end{itemize}

%%%-------------------------------- A Key Corollary
%Consider any maximal independent set $T=T_1\times T_2\times \dots \times T_t\in \Ic_{\rm max}(\Gamma_t)$. By Lemma \ref{lem:indep} we know that $T_j\in \Ic_{\rm max}(\Gamma)$, $\forall j\in [t]$. We have the following corollary. 
%
%\begin{corollary}  \label{cor:associated:hypergraph:product}
%%Consider any maximal independent set $T=T_1\times T_2\times \dots \times T_t\in \Ic_{\rm max}(\Gamma_t)$. Then, we know 
%We have $\Hc(T)=\Hc(T_1)\times \Hc(T_2) \times \dots \times \Hc(T_t)$, where for any two hypergraphs $\Hc_1$ and $\Hc_2$, the product $\Hc_1\times \Hc_2$ is defined in \cite[Section 1.6]{scheinerman2011fractional}. Moreover, we have $k_f(\Hc(T))=\prod_{j\in [t]}k_f(\Hc(T_j))$. 
%\end{corollary}
%
%The proof of the above corollary follows immediately from Lemma \ref{lem:intersection:indep:neighbour} and \cite[Theorem 1.6.1]{scheinerman2011fractional}. 

%We show $\lim_{t\to \infty}\frac{1}{t} \log (1-(1-(\frac{1}{k_f})^t)^{k(t)})\ge \log \frac{1}{k_f}+\sigma$ below. First, for any positive real number $m< 2^{\sigma}$, we have 
Next, we have 
\begin{align}
&\lim_{t\to \infty}\frac{1}{t} \log (1-(1-(\frac{1}{k_f})^t)^{g(t)})  \nonumber  \\
&\stackrel{(j)}{>} \lim_{t\to \infty}\frac{1}{t} \log (1-(1-(\frac{1}{k_f})^t)^{m^t})  \nonumber  \\
&=\lim_{t\to \infty}\frac{1}{t} \log \Big( (1-(1-(\frac{1}{k_f})^t)) (\sum_{j\in [0:m^t-1]}(1-(\frac{1}{k_f})^t)^j) \Big)  \nonumber  \\
&=\log \frac{1}{k_f} + \lim_{t\to \infty}\frac{1}{t} \log \sum_{j\in [0:m^t-1]}(1-(\frac{1}{k_f})^t)^j  \nonumber  \\
&\ge \log \frac{1}{k_f} + \lim_{t\to \infty}\frac{1}{t} \log (m^t\cdot (1-(\frac{1}{k_f})^t)^{m^t-1})  \nonumber  \\
&=\log \frac{1}{k_f} + \log m + \lim_{t\to \infty} \frac{(m^t-1)\log (1-(\frac{1}{k_f})^t)}{t}  \nonumber  \\
&\stackrel{(k)}{=} \log \frac{1}{k_f} + \log m,  \label{eq:calculus:result}
\end{align}
where (j) follows from \eqref{eq:multiple:approximate:guesses:m:1}, 
and (k) follows from the fact that $\lim_{t\to \infty} \frac{(m^t-1)\log (1-(\frac{1}{k_f})^t)}{t}=0$ , which in turn follows from \eqref{eq:multiple:approximate:guesses:m:2}. 

Finally, as \eqref{eq:calculus:result} holds for any positive $m< 2^{\sigma}$, we have  
\begin{align}
\lim_{t\to \infty}\frac{1}{t} \log (1-(1-(\frac{1}{k_f})^t)^{g(t)})\ge \log \frac{1}{k_f} + \sigma.  \label{eq:multiple:approximate:guesses:limits}
\end{align}

Combining \eqref{eq:multiple:approximate:guesses:mid} and \eqref{eq:multiple:approximate:guesses:limits}, we have
\begin{align}
\Lc_{\Theta,g}
&=\lim_{t\to \infty} \frac{1}{t} \inf_{\substack{P_{Y|X^t}: \\ \Xc^t(y)\in \Ic(\Gamma_t),\forall y\in \Yc}}  \nonumber  \\
&\qquad \quad \log \max_{P_X} \frac{\sum\limits_{y\in \Yc} \max\limits_{\substack{K\subseteq \Xc^t:\\|K|\le g(t)}} \sum\limits_{x^t\in N(\Theta_t,K)} P_{X^t,Y}(x^t,y)}{\max\limits_{\substack{K\subseteq \Xc^t:|K|\le g(t)}} \sum\limits_{x^t\in N(\Theta_t,K)} P_{X^t}(x^t)}  \nonumber  \\
&\ge  \lim_{t\to \infty}\frac{1}{t}\log \max_{P_X} \frac{\sum_{y\in \Yc} (1-(1-(\frac{1}{k_f})^t)^{g(t)})P_Y(y)}{\max\limits_{\substack{K\subseteq \Xc^t:|K|\le g(t)}} \sum\limits_{x^t\in N(\Theta_t,K)} P_{X^t}(x^t)}  \nonumber  \\
&=\lim_{t\to \infty}\frac{1}{t}\log \max_{P_X} \frac{1-(1-(\frac{1}{k_f})^t)^{g(t)}}{\max\limits_{\substack{K\subseteq \Xc^t:|K|\le g(t)}} \sum\limits_{x^t\in N(\Theta_t,K)} P_{X^t}(x^t)}  \nonumber  \\
&\ge \lim_{t\to \infty}\frac{1}{t}\log \max_{P_X} \frac{1-(1-(\frac{1}{k_f})^t)^{g(t)}}{g(t)\cdot (\max_{x\in \Xc}P_X(N(\Theta,x)))^t}  \nonumber  \\
&=\log \frac{1}{\min\limits_{P_X} \max\limits_{x\in \Xc}P_X(N(\Theta,x))}+\lim_{t\to \infty}\frac{1}{t}\log \frac{1}{g(t)}+  \nonumber  \\
&\qquad \lim_{t\to \infty}\frac{1}{t} \log (1-(1-(\frac{1}{k_f})^t)^{g(t)})  \nonumber  \\
&\ge \log \frac{1}{\min\limits_{P_X} \max\limits_{x\in \Xc}P_X(N(\Theta,x))}-\sigma+\log \frac{1}{k_f}+\sigma  \nonumber  \\
&=\log \frac{1}{\min\limits_{P_X} \max\limits_{x\in \Xc}P_X(N(\Theta,x))}+\log \frac{1}{k_f}  \nonumber  \\
&=\log \frac{p_f(\Theta)}{k_f}
\end{align}
\Rarxiv{where the last equality follows from the fact that $p_f(\Theta)=\frac{1}{\min_{P_X} \max_{x\in \Xc}P_X(N(\Theta,x))}$, which has been proved towards the end of Appendix \ref{app:proof:thm:one:approximate:guess}}. 
%which completes the proof of the theorem.  
\end{IEEEproof}